\documentstyle[prd,eqsecnum,aps]{revtex}
\def\mib#1{\mbox{\boldmath $#1$}}
\def\lf{\left}\def\ri{\right}
\tighten

\begin{document}
 \title{Operator relation among neutrino fields and oscillation formulas in matter}
\author{Kanji  Fujii$^{*1}$, Chikage Habe$^{*1}$
 and Massimo Blasone$^{*2}$
\\ [2mm]
{\it
\begin{tabular}{c}
$^{*1}$ Department of Physics, Faculty of Science,
Hokkaido University, Sapporo 060-0810, Japan
\\ [2mm]
$^{*2}$ Blackett  Laboratory, Imperial  College,   Prince Consort Road,
London  SW7 2BW, U.K.
\\ and Dipartimento di  Fisica and INFN,
Universit\`a di Salerno, I-84100 Salerno, Italy
\end{tabular}
}}
 \date{\today}
 \maketitle
 \begin{abstract}
Starting with the relation between two kinds of neutrino field
operators corresponding to  definite flavor  and mass states, we
investigate if  the usual evolution relations for neutrino
oscillation in matter can be consistently derived, in the case of
high energy neutrinos. A negative answer is obtained contrary to
an earlier result [P.~D.~Mannheim, Phys.\ Rev.\ D\ {\bf 37} (1988)
1935]. The reason for such a difference relies essentially on the
Bogolyubov transformation between two kinds of momentum-helicity
creation- and annihilation-operators with definite flavors and
masses, which was not taken into account in previous treatments.
\end{abstract}

\section{Introduction}\label{sec_introduction}
In view of the great phenomenological and theoretical relevance of
neutrino mixing and oscillations,  much effort has been recently
devoted to a full quantum field theoretical  analysis of this phenomenon both
in vacuum as well as in matter\cite{MAN,GKL,GS,BV1,FHY1,BV2,BHV,CARD,HL,FHY2,BCV,Ji,YI,BEUTHE}.
Among the various approaches which have been developed, the one proposed by Blasone and
Vitiello\cite{BV1}(hereafter, referred to BV) is based on two
kinds of Hilbert spaces, i.e. the flavor- and the
mass-Hilbert spaces, and related considerations in the case
of vacuum oscillations have been published in Refs.\cite{BV1,FHY1,BV2,BHV,HL,FHY2,BCV}.
In these references it has been pointed out that
non-standard oscillation terms due to the
non-trivial nature of the mixing transformations
do indeed appear in the full  oscillation formula obtained in quantum field theory.

The aim of the present work is to develop field-theoretical
considerations concerning neutrino oscillations in matter along
the line of BV\cite{BV1} and the one developed by Fujii, Habe and Yabuki\cite{FHY1}.
We examine this task by assuming the static medium and the
extreme-relativistic case which are to be defined exactly  in Sec.II.
In this respect,  it is useful to remember  that, when we want to define a
flavor neutrino state with definite momentum and helicity, there
appear arbitrary mass parameters $\mu_{\lambda}$'s\cite{FHY1,BV2,FHY2},
so that we cannot estimate the magnitude of  $\mu _{\lambda}$'s in
comparison with the neutrino momentum.
Mannheim has investigated the same problem  in the paper\cite{MAN}
published in somewhat early stage\footnote{When one of
the present authors (K.F.) reported the main part of
the present work at International Workshop on the Quantum Field Theory of
Particle Mixing and Oscillations, 13-15 June 2002, Vietri sul Mare (Salerno),
Dr. C.Y. Cardall called attention to Mannheim's work.}:
his conclusion is that the standard Mikheyev-Smirnov-Wolfenstein (MSW)
formalism\cite{MSW} is derived.
However, after the considerations developed in the following, we obtain a negative
conclusion against that of Mannheim\cite{MAN}.
The essential point is that, in quantities
related to the evolution relation of flavor neutrino states, Mannheim
dropped the effect coming from Bogolyubov transformation between the two
kinds of creation- and annihilation-operators (with definite momentum and
helicity) corresponding to the flavor- and the mass-eigenstates.
We discuss this point in detail in this paper.

In the following we consider the problem in the 2-flavor case for simplicity,
although the above conclusion will be seen to hold true also in the many flavor case.

\section{Preliminaries}\label{Prelim}
\subsection{General framework}\label{framework}
We investigate the neutrino oscillation in matter on the basis
 of the electro-weak unified theory, in which the known
 neutrinos participate in the weak interaction of V-A type, mediated by
 $W^{\pm}$- and $Z^{0}$-bosons.
For low momentum-transfers, $|q^{2}|\ll m^{2}_{W}, m^{2}_{Z}$,
the effective interaction Lagrangian is
\begin{eqnarray}
L^{eff}_{weak}(x)=&-&\frac{G_{F}}{\sqrt{2}}\Big\{(\bar\nu _{e}v^{\alpha}e)
({\bar e}v_{\alpha}\nu_{e})+(\bar\nu _{\mu}v^{\alpha}\mu)({\bar \mu}
v_{\alpha}\nu_{\mu})+\cdots \nonumber \\
&+&\frac{\kappa _{\omega}}{2}(\sum_{\rho}\bar\nu _{\rho}v^{\alpha}\nu_{\rho})
\left(-\bar e v_{\alpha}e+4\sin^{2}\theta_{W}\bar e \gamma_{\alpha}e\right)+\cdots
\Big\}\nonumber \\
&=&-\frac{G_{F}}{\sqrt{2}}\Big\{\bar\nu _{e}v^{\alpha}\nu_{e}
((1-\frac{\kappa_{\omega}}{2})\bar e v_{\alpha}e+2\kappa_{\omega}\sin^{2}\theta_{W}
\bar e \gamma_{\alpha}e)
\nonumber\\
&+&\sum_{\rho\neq e}(\bar\nu _{\rho}v^{\alpha}\nu_{\rho})\kappa_{\omega}
(-\frac{1}{2}\bar e v_{\alpha}e+2\sin^{2}\theta_{W}\bar e \gamma_{\alpha}e)+\cdots\Big\},
\end{eqnarray}
where $v^{\alpha}=\gamma^{\alpha}(1+\gamma_{5})$ with $\vec\gamma=-\rho _{y}
\otimes\vec\sigma, \gamma^{4}=\rho _{x}\otimes I, \gamma _{5}=-\rho_{z}\otimes I$
in the Kramers representation\cite{CORN,FHY1,FHY2};
$\sqrt{\kappa_{w}}{m_{z}}/{m_{w}}=1/\cos\theta_{w}, (\kappa_{w})_{exp}\approx 1;
G_{F}\approx 1.027\times 10^{-5}/m^{2}_{P}$.
Hereafter we take $\kappa_{w}=1$.
(As to the details of the notations below, see Refs.\cite{FHY1,FHY2}.)

For neutrino propagation in matter, we define
\begin{eqnarray}
n^{(e)}(x)_{\alpha}=i\langle\bar e(x) v_{\alpha}e(x)\rangle _{matt}\approx
i\langle\bar e(x) \gamma_{\alpha}e(x)\rangle _{matt},
\end{eqnarray}
which plays the role of an external field.
Under the adiabatic approximation, $(n^{(e)}(x)_{\alpha})$ may be regarded
as $x$-independent over a wide range.
With the aim of presenting the logical structure of the problem as simply
as possible, we assume $(n^{(e)}(x)_{\alpha})$ to be $x$-independent
in the same way as Mannheim\cite{MAN}.
We define the time-like 4-vector $(u_{\alpha})=(\vec u, iu_{0})$ with
$u^{\alpha}u_{\alpha}=-1$ such that it  reduces
to $(\vec 0, i)$ in the rest frame of medium;
thus, we can write $(n^{(e)}(x)_{\alpha})=(n_{(e)}u_{\alpha})$ which
reduces to $(\vec 0, i n_{(e)})$ in the rest frame of medium.

By noting that
\begin{eqnarray}
\frac{1}{i}\not n ^{(e)}P_{L}=\not n ^{(e)}\frac{1+\gamma _{5}}{2i}=n_{(e)}
\left(\begin{array}{cc}
0&\vec\sigma\vec u+u_{0}I\\0&0
\end{array}\right),
\end{eqnarray}
and by defining
\begin{eqnarray}
b_{1}\equiv \sqrt{2}G_{F}n_{(e)}\quad, \quad
b_{0}\equiv \sqrt{2}G_{F}n_{(e)}(2\sin^{2}\theta_{w}-\frac{1}{2}),
\end{eqnarray}
the effective Lagrangian related with the present consideration for the 2-flavor case in expressed as
\begin{eqnarray}\label{2.4}
L^{matt}_{eff}(x)&=&-
\left(\begin{array}{cc}
\bar\nu_{e}(x)&\bar\nu_{\mu}(x)
\end{array}\right)
\lf[\not\partial +M^{0}+
\left(\begin{array}{cc}
(b_{0}+b_{1})\not u P_{L}/i&0\\
0&b_{0}\not u P_{L}/i
\end{array}\right)
\ri]
\left(\begin{array}{c}
\nu _{e}(x)\\ \nu _{\mu}(x)
\end{array}\right)+L_{int}(x)\nonumber\\
&=&-\left(\begin{array}{cc}
\bar\nu_{e}(x)&\bar\nu_{\mu}(x)
\end{array}\right)
\lf[\not\partial +
\left(\begin{array}{cccc}
m_{ee}&N&m_{e\mu}&0\\
0&m_{ee}&0&m_{e\mu}\\
m_{\mu e}&0&m_{\mu\mu}&N'\\
0&m_{\mu e}&0&m_{\mu\mu}
\end{array}\right)
\ri]
\left(\begin{array}{c}
\nu _{e}(x)\\ \nu _{\mu}(x)
\end{array}\right)+L_{int}(x)\nonumber\\
&\equiv&L^{matt}_{F}(x)+L_{int}(x),
\end{eqnarray}
where $L_{int}(x)$ is assumed to include no bilinear term of neutrino
fields, and hereafter its contributions are neglected;
$N$ and $N'$ are defined by
\begin{eqnarray}\label{2.5}
N\equiv (b_{0}+b_{1})(\vec\sigma\vec u+u_{0}I)\quad, \quad N'=b_{0}(\vec\sigma\vec u+u_{0}I).
\end{eqnarray}

The mass matrix $M^{0}$ in the 2-flavor case can be taken with no loss
of generality, as $m_{e\mu}=m_{\mu e}, m_{\mu\mu}>m_{ee}\geq 0$;
then, the eigenvalues, $m_{1}$ and $m_{2}$,
\begin{eqnarray}
m_{1(2)}=\frac{1}{2}\left( m_{ee}+m_{\mu\mu}-(+)\sqrt{(m_{\mu\mu}-m_{ee})^{2}+4m_{e\mu}^{2}}\right)
\end{eqnarray}
satisfy
\begin{eqnarray}
m_{1}&\geq&0\quad\mbox{ for } \quad \sqrt{m_{ee}m_{\mu\mu}} \geq |m_{e\mu}|,\nonumber\\
m_{1}&<&0\quad \mbox{ for }  \quad \sqrt{m_{ee}m_{\mu\mu}} < |m_{e\mu}|.
\end{eqnarray}
From the mixing relation of the "mass-eigenfield" $\nu_{j}(x)$'s
\begin{eqnarray}
\nu_{\rho}(x)=\sum _{j}z^{1/2}_{\rho j}\nu _{j}(x)\quad
\mbox{ with } \quad Z^{1/2}=[z^{1/2}_{\rho j}]
\end{eqnarray}
satisfying $Z^{1/2\dagger}Z^{1/2}=I$ as well as
\begin{eqnarray}
Z^{1/2\dagger}M^{0}Z^{1/2}=m_{diag}=
\left(\begin{array}{cc}
m_{1}&0\\0&m_{2}
\end{array}\right),
\end{eqnarray}
we can take
\begin{eqnarray}
\left(\begin{array}{c}
\nu_{e}(x)\\\nu_{\mu}(x)
\end{array}\right)
=
\left(\begin{array}{cc}
\cos\theta&\sin\theta\\-\sin\theta&\cos\theta
\end{array}\right)
\left(\begin{array}{c}
\nu_{1}(x)\\\nu_{2}(x)
\end{array}\right)
\end{eqnarray}
with
\begin{eqnarray}
\tan\theta =\frac{1}{2m_{e\mu}}[-(m_{\mu\mu}-m_{ee})+\sqrt{(m_{\mu\mu}-m_{ee})^{2}+4m_{e\mu}^{2}}].
\end{eqnarray}
For later convenience we write the relations among $m_{\rho\sigma}$'s, $m_{j}$'s and $\theta$;
\begin{eqnarray}
\left(\begin{array}{c}
m_{ee}\\m_{\mu\mu}
\end{array}\right)
&=&
\left(\begin{array}{cc}
\cos^{2}\theta&\sin^{2}\theta\\\sin^{2}\theta&\cos^{2}\theta
\end{array}\right)
\left(\begin{array}{c}
m_{1}\\m_{2}
\end{array}\right),\\
m_{e\mu}&=&(m_{2}-m_{1})\sin\theta \cos\theta \quad ,\quad
\tan(2\theta)=\frac{2m_{e\mu}}{m_{\mu\mu}-m_{ee}}.
\end{eqnarray}

Employing the above relations, we can write $L^{matt}_{(F)}(x)$ given
by (\ref{2.4}) in terms of the mass-basis operators;
\begin{eqnarray}\label{2.10a}
L^{matt}_{(F)}(x)
&=&-
\left(\begin{array}{cc}
\bar\nu _{1}(x)&\bar\nu _{2}(x)
\end{array}\right)
\lf[\not\partial +
\left(\begin{array}{cccc}
m_{1}&B&0&G\\
0&m_{1}&0&0\\
0&G&m_{2}&B'\\
0&0&0&m_{2}
\end{array}\right)
\ri]
\left(\begin{array}{c}
\nu _{1}(x)\\\nu _{2}(x)
\end{array}\right)\nonumber\\
&\equiv&L^{matt}_{(M)}(x),
\end{eqnarray}
where $2\times 2$ matrices $B$, $B'$ and $G$ are defined by
\begin{eqnarray}\label{2.16'}
\left(\begin{array}{cc}
B&G\\G&B'\\
\end{array}\right)
=\kappa\otimes (\sigma\vec u+u_{0}I)
\end{eqnarray}
with $\kappa _{11}=b_{0}+b_{1}\cos^{2}\theta$ ,
$\kappa _{12}=\kappa _{21}=b_{1}\cos\theta \sin\theta$ and  $\kappa _{22}=b_{0}+b_{1}\sin^{2}\theta$.

\subsection{Effective Hamiltonian and eigenvalue equation}\label{EffH}
The effective Hamiltonian obtained from
$L^{matt}_{(M)}(x)$, (\ref{2.10a}), is
\begin{eqnarray}
H^{matt}_{(M)}(x)
&=&:
\left(\begin{array}{cc}
\bar\nu _{1}(x)&\bar\nu _{2}(x)
\end{array}\right)
\lf[\vec\gamma \vec\nabla +
\left(\begin{array}{cccc}
m_{1}&B&0&G\\
0&m_{1}&0&0\\
0&G&m_{2}&B'\\
0&0&0&m_{2}
\end{array}\right)\ri]
\left(\begin{array}{c}
\nu _{1}(x)\\\nu _{2}(x)
\end{array}\right):\nonumber\\
&=&:
\left(\begin{array}{cc}
\nu _{1}^{\dagger}(x)&\nu _{2}^{\dagger}(x)
\end{array}\right)
\left(\begin{array}{cccc}
-i\vec\sigma\vec\nabla&m_{1}&0&0\\
m_{1}&i\vec\sigma\vec\nabla+B&0&G\\
0&0&-i\vec\sigma\vec\nabla&m_{2}\\
0&G&m_{2}&i\vec\sigma\vec\nabla+B'
\end{array}\right)
\left(\begin{array}{c}
\nu _{1}(x)\\\nu _{2}(x)
\end{array}\right):.
\end{eqnarray}
We expand the $\nu_{j}(x)$ fields ($j=1, 2$) as
\begin{eqnarray}\label{2.12}
\nu_{j}(x)&=&\frac{1}{\sqrt{V}}\sum_{\vec k r}
\lf\{\alpha _{j}(kr;x^{0})u_{j}(kr)e^{i\vec k\vec x}+
\beta^{\dagger} _{j}(kr;x^{0})v_{j}(kr)e^{-i\vec k\vec x}\ri\}\nonumber
\\
&=&\frac{1}{\sqrt{V}}\sum_{\vec k r}\lf\{\alpha _{j}(kr;x^{0})u_{j}(kr)+
\beta^{\dagger} _{j}(-kr;x^{0})v_{j}(-kr)e^{i\vec k\vec x}\ri\};
\end{eqnarray}
here, $u_{j}(kr)$ and $v_{j}(kr)$ are the momentum-helicity eigenfunctions satisfying
\begin{eqnarray}
(i\not k+m_{j})u_{j}(kr)=0\,,\quad (-i\not k+m_{j})v_{j}(kr)=0 \quad
\mbox{ with }k_{0}=\sqrt{\vec k^{2}+m_{j}^{2}}\equiv \omega_{j}(k),\\
\beta(-kr;x^{0})\equiv\beta(-\vec k,k_{0},r;x^{0})\,,\quad  v(-kr)\equiv v(-\vec k,k_{0},r).\label{2.13b}
\end{eqnarray}
By employing the concrete forms of $u_{j}(kr)$ and $v_{j}(-kr)$
given in Appendix A, we can express the Hamiltonian
\begin{eqnarray}
{\cal H}_{(M)}^{matt}(x^{0})\equiv \int d^{3}x H^{matt}_{(M)}(x),
\end{eqnarray}
in terms of the expansion-coefficient operators in
(\ref{2.12}).
(See Table I as well as Eqs.(\ref{A9}) and (\ref{A10}) in Appendix A.)
We obtain
\begin{eqnarray}\label{2.15}
{\cal H}_{(M)}^{matt}(x^{0})=\sum _{\vec k r}\sum _{j}:
w_{j}(k)(\alpha^{\dagger}_{j}(kr)\alpha _{j}(kr)-\beta_{j}(-kr)
\beta^{\dagger} _{j}(-kr)):+\int d^{3}x\sum_{j,l}\kappa _{jl}:\nu^{\dagger}_{j}(x)
\left(\begin{array}{cc}
0&0\\0&\vec\sigma\vec u+u_{0}
\end{array}\right)
\nu _{l}(x):
\end{eqnarray}
This shows that, due to the non-static($\vec u\neq 0$) contribution,
${\cal H}^{matt}_{(M)}(x^{0})$ is not separable into two parts,
corresponding to helicity $r=\uparrow$ or $\downarrow$,
since the last integral part of (\ref{2.15}) is equal to
\begin{eqnarray}\label{2.16}
\sum _{j,l}\kappa _{jl}:\int d^{3}x (u_{0}\nu^{\dagger}_{j}(x)P_{L}
\nu _{l}(x)+\nu^{\dagger}_{j}(x)(\vec\sigma\vec u)P_{L}\nu _{l}(x)    ):.
\end{eqnarray}
In the static-medium approximation, i.e. $(\vec u,\vec u_{0})
\rightarrow(\vec 0,1)$, ${\cal H}^{matt}_{(M)}(x^{0})$ is separable into two parts;
in the two-flavor case we are lead to consider the $4\times 4$
matrices ${\cal H}^{matt}_{(M)}(k\uparrow)$ and ${\cal H}^{matt}_{(M)}(k\downarrow)$ as given below.

From Eq.(\ref{2.15}), we obtain the Hamiltonian in the static-medium case
\begin{eqnarray}\label{2.17a}
{\cal H}_{(M)}^{matt}(x^{0})|_{static}&=&\sum _{\vec k,r}{\cal H}_{(M)}^{matt}(kr;x^{0}),\\
{\cal H}_{(M)}^{matt}(kr;x^{0})&=&:
\left(\begin{array}{cc}
{\mib\alpha}^{\dagger}_{M}(kr;x^{0})&{\mib\beta}_{M}(-kr;x^{0})\\
\end{array}\right)
H_{(M)}^{matt}(kr)
\left(\begin{array}{c}
{\mib\alpha}_{M}(kr;x^{0})\\{\mib\beta}^{\dagger}_{M}(-kr;x^{0})\\
\end{array}\right):\label{2.17b}
\end{eqnarray}
with
\begin{eqnarray}\label{2.17c}
H_{(M)}^{matt}(k\uparrow)&=&
\left(\begin{array}{cccc}
\omega _{1}+\kappa _{11}s_{1}s_{1}&\kappa _{12}s_{1}s_{2}
&i\kappa _{11}s_{1}c_{1}&i\kappa _{12}s_{1}c_{2}\\
\kappa _{12}s_{2}s_{1}&\omega _{2}+\kappa _{22}s_{2}s_{2}
&i\kappa _{12}s_{2}c_{1}&i\kappa _{22}s_{2}c_{2}\\
-i\kappa _{11}c_{1}s_{1}&-i\kappa _{12}c_{1}s_{2}
&-\omega _{1}+\kappa _{11}c_{1}c_{1}&\kappa _{12}c_{1}c_{2}\\
-i\kappa _{12}c_{2}s_{1}&-i\kappa _{22}c_{2}s_{2}
&\kappa _{12}c_{2}c_{1}&-\omega _{2}+\kappa _{22}c_{2}c_{2}
\end{array}\right),\\
H_{(M)}^{matt}(k\downarrow)&=&
\left(\begin{array}{cccc}
\omega _{1}+\kappa _{11}c_{1}c_{1}&\kappa _{12}c_{1}c_{2}
&-i\kappa _{11}c_{1}s_{1}&-i\kappa _{12}c_{1}s_{2}\\
\kappa _{12}c_{2}c_{1}&\omega _{2}+\kappa _{22}c_{2}c_{2}
&-i\kappa _{12}c_{2}s_{1}&-i\kappa _{22}c_{2}s_{2}\\
i\kappa _{11}s_{1}c_{1}&i\kappa _{12}s_{1}c_{2}
&-\omega _{1}+\kappa _{11}s_{1}s_{1}&\kappa _{12}s_{1}s_{2}\\
i\kappa _{12}s_{2}c_{1}&i\kappa _{22}s_{2}c_{2}
&\kappa _{12}s_{2}s_{1}&-\omega _{2}+\kappa _{22}s_{2}s_{2}
\end{array}\right).\label{2.17d}
\end{eqnarray}
Here we use the notation (in the two-flavor case)
\begin{eqnarray}
{\mib\alpha} _{M}(kr;x^{0})=
\left(\begin{array}{c}
\alpha _{1}(kr;x^{0})\\\alpha _{2}(kr;x^{0})
\end{array}\right)\quad ,
\quad  {\mib\beta} _{M}(-kr;x^{0})=
\left(\begin{array}{c}
\beta _{1}(-kr;x^{0})\\\beta _{2}(-kr;x^{0})
\end{array}\right);\\
c_{j}=\cos(\chi _{j}/2)\quad , \quad
s_{j}=\sin(\chi _{j}/2),\,\,  cot\chi _{j}=|\vec k|/m_{j}.\nonumber
\end{eqnarray}

The equations determining the energy eigenvalues are
\begin{eqnarray}\label{2.18}
\det[H^{matt}_{(M)}(kr)]=0\;, \qquad  r=\uparrow, \downarrow.
\end{eqnarray}
The concrete forms of LHS of (\ref{2.18}) are given by
\begin{eqnarray}\label{2.19a}
\det[H^{matt}_{(M)}(k\uparrow)-\lambda]=(\lambda^{2}-\omega^{2}_{1})
(\lambda^{2}-\omega^{2}_{2})-(\lambda-k)\{\kappa _{11}(\lambda^{2}-
\omega^{2}_{2})+\kappa _{22}(\lambda^{2}-\omega^{2}_{1}) \}+(\lambda-k)^{2}\det\kappa,\\
\det[H^{matt}_{(M)}(k\downarrow)-\lambda]=(\lambda^{2}-\omega^{2}_{1})
(\lambda^{2}-\omega^{2}_{2})-(\lambda+k)\{\kappa _{11}(\lambda^{2}-
\omega^{2}_{2})+\kappa _{22}(\lambda^{2}-\omega^{2}_{1}) \}+(\lambda+k)^{2}\det\kappa.
\label{2.19b}
\end{eqnarray}
Hereafter we write $|\vec k|$ simply as $k$ when there is no fear of confusion.

The relation of the eigenvalues obtained from Eq.(\ref{2.18}) with the poles
of flavor neutrino propagators will be examined in Sec.III.
Here we discuss  the relation of Eq.(\ref{2.18}) with the  eigenvalue
equation obtained  by Mannheim \cite{MAN}, i.e. Eq.(2.16) in his paper:
\begin{eqnarray}\label{2.20}
E^{4}-E^{3}V&-&E^{2}(2k^{2}+Vk+m^{2}_{1}+m^{2}_{2})+EV(k^{2}+
m^{2}_{1}\sin^{2}\theta+m^{2}_{2}\cos^{2}\theta)\nonumber\\
&+&\omega^{2}_{1}\cdot\omega^{2}_{2}+Vk(k^{2}+m^{2}_{1}\sin^{2}\theta+
m^{2}_{2}\cos^{2}\theta)=0;
\end{eqnarray}
$V=\sqrt{2}G_{F}n_{(e)}=b_{1}$ in our notation.
In (\ref{2.20}) the neutral current contribution is subtracted
from $H^{matt}_{(M)}(k\downarrow)$ as a common term, so that in our case we have to take
\begin{eqnarray}
\kappa\longrightarrow b_{1}
\left(\begin{array}{cc}
\cos^{2}\theta&\cos\theta \sin\theta\\\cos\theta \sin\theta&\sin^{2}\theta
\end{array}\right), \quad \det\kappa\longrightarrow 0.
\end{eqnarray}
Then, we obtain from (\ref{2.19b})
\begin{eqnarray}
\det[H^{matt}_{(M)}(k\downarrow)-\lambda]\longrightarrow(E^{2}-
\omega^{2}_{1})(E^{2}-\omega^{2}_{2})-(E+k)V(E^{2}-k^{2}-
m^{2}_{1}\sin^{2}\theta-m^{2}_{2}\cos^{2}\theta),
\end{eqnarray}
which coincides with LHS of (\ref{2.20}).

In Ref.\cite{MAN}, Mannheim pointed out that, in the limit
of high energies and weak potentials, the standard
MSW formalism for neutrino
oscillations in matter\cite{MSW} is derived on the basis of the field theory.
It seems necessary, however, for us to reexamine this conclusion
 in detail because of the following reasons.

Hereafter we use the terminology ``{\it extreme-relativistic case}''
to denote the limit of higher neutrino-energies and weak potentials, i.e.
\begin{eqnarray}\label{2.23}
|\vec k|^{2} \gg \{m_{j}^{2}\mbox{'s}, A=2kb_{1}, A_{0}=2kb_{0} \}.
\end{eqnarray}
Since the mass eigenvalues are obtained by taking $\omega _{j}\simeq
k+\frac{m_{j}^{2}}{2k}$, we use
\begin{eqnarray}\label{2.24}
c_{j}\simeq 1-\frac{m^{2}_{j}}{8k^{2}},\quad s_{j}\simeq \frac{m_{j}}{2k}.
\end{eqnarray}
If we simply take the {\it super-relativistic limit}
\begin{eqnarray}\label{2.25a}
c_{j}\longrightarrow 1,\quad   s_{j}\longrightarrow 0 \quad \forall j,
\end{eqnarray}
then Eq.(\ref{2.17d}) reduces to
\begin{eqnarray}\label{2.25b}
H^{matt}_{(M)}(k\downarrow)\longrightarrow
\left(\begin{array}{cccc}
\omega _{1}+\kappa _{11}&\kappa _{12}&0&0\\
\kappa _{12}&\omega _{2}+\kappa _{22}&0&0\\
0&0&-\omega _{1}&0\\
0&0&0&-\omega _{2}
\end{array}\right)
\end{eqnarray}
which is separable with respect to ${\mib\alpha}_{(M)}(k\downarrow ;x^{0})$
and ${\mib\beta}^{\dagger}_{(M)}(k\downarrow ;x^{0})$, and leads
to the usually employed Hamiltonian appearing in Schr\"odinger
equation for one-particle states of two-flavor neutrinos.
(See Eq.(2.23) in Ref.\cite{MAN} and the next subsection.)
The eigenvalue equation for (\ref{2.25b}) is given by
\begin{eqnarray}\label{2.26}
\det[H^{matt}_{(M)}(k\downarrow)|_{c_{j}=1, s_{j}=0}-\lambda]=
(\omega _{1}+\lambda)(\omega _{2}+\lambda)\lf\{(\omega _{1}-\lambda)
(\omega _{2}-\lambda)+(\omega _{1}-\lambda)\kappa _{22}+
(\omega _{2}-\lambda)\kappa _{11}+\det\kappa\ri\}.
\end{eqnarray}
The curly bracket part in RHS of (\ref{2.26}) and the remaining part determine the eigenvalues of positive- and negative-energy modes, respectively.
If we take $\omega_{j}, j=1,2$ to be $k$, which is consistent with (\ref{2.25a}), the curly bracket part of (\ref{2.26}) is 
\begin{eqnarray}\label{}
\{\cdots \} \mbox{ of (\ref{2.26})}=(k-\lambda)(k-\lambda)+(k-\lambda)(\kappa _{11}+\kappa _{22})+\det\kappa ,
\end{eqnarray}
leading to the eigenvalues
\begin{eqnarray}\label{}
\lambda-k &=&\frac{1}{2}\left[(\kappa _{11}+\kappa _{22})\pm \sqrt{(\kappa _{11}-\kappa _{22})^{2}+4(\kappa_{12})^{2}}\right]\nonumber\\
&=&\frac{1}{4k}\left[2A_{0}+A\pm A\right].
\end{eqnarray}
If we write as usual
\begin{eqnarray}\label{}
\lambda=k+\frac{\tilde m^{2}}{2k},
\end{eqnarray}
we obtain
\begin{eqnarray}\label{}
\tilde m^{2}=A_{0}+A \mbox{ or } A_{0},
\end{eqnarray}
which are different from the usual ones\cite{MAN,FY} apart from the common factor $A_{0}$.

In order to obtain the mass eigenvalues $\{\tilde m^{2}$'s$\}$, we have to examine the "exact" expression (\ref{2.19b}), i.e.
\begin{eqnarray}\label{2.27}
\det[H^{matt}_{(M)}(k\downarrow)-\lambda] &=&(\omega
_{1}+\lambda)(\omega _{2}+\lambda)\Big\{(\omega _{1}-\lambda)
(\omega _{2}-\lambda)
\nonumber\\
&+&\frac{k+\lambda}{\omega _{1}+\lambda} (\omega
_{2}-\lambda)\kappa _{11}+\frac{k+\lambda}{\omega _{2}+
\lambda}(\omega _{1}-\lambda)\kappa
_{22}+\frac{(k+\lambda)^{2}}{(\omega _{1}+ \lambda)(\omega
_{2}+\lambda)}\det\kappa  \Big\}.
\end{eqnarray}
When we take, in the approximation of the extreme-relativistic case,
\begin{eqnarray}\label{2.45'}
\omega_{j}\simeq k+\frac{m_{j}^{2}}{2k} \mbox{ and } \lambda= k+\frac{\tilde m^{2}}{2k},
\end{eqnarray}
we obtain
\begin{eqnarray}\label{2.29b}
\{\cdots \}\mbox{-part in (\ref{2.27})}&\simeq& \frac{1}{4k^{2}}(m^{2}_{1}-\tilde m^{2})(m^{2}_{2}-\tilde m^{2})+
\lf(1-\frac{m^{2}_{1}}{4k^{2}}\ri)
\frac{k\kappa _{11}}{k}\cdot\frac{m^{2}_{2}-\tilde m^{2}}{2k}+
\lf(1-\frac{m^{2}_{2}}{4k^{2}}\ri)\frac{k\kappa _{22}}{k}\cdot\frac{m^{2}_{1}-
\tilde m^{2}}{2k}\nonumber\\
&&+\lf(1-\frac{m^{2}_{1}+m^{2}_{2}}{4k^{2}}\ri)\frac{k^{2}\det\kappa}{k^{2}}.
\end{eqnarray}
In accordance with the approximation in the extreme-relativistic case, we neglect all terms with higher order than $\{m^{2}_{j},\tilde m^{2},A,A_{0}\}/k^{2}$ in comparison with 1, and correction terms to terms such as (relevant mass)$^{2}$ and $k\kappa_{ij}$ are to be dropped.
Then (\ref{2.29b}) is consistently expressed as
\begin{eqnarray}\nonumber
\{\cdots \}\mbox{-part in (\ref{2.27})}&\simeq&
\frac{1}{4k^{2}}(m^{2}_{1}-\tilde m^{2})(m^{2}_{2}-\tilde
m^{2})+\frac{A_{0}+A\cos^{2}\theta}{2k}\cdot\frac{m^{2}_{2}-\tilde
m^{2}}{2k}+\frac{A_{0}+A\sin^{2}\theta}{2k}\cdot\frac{m^{2}_{1}-\tilde
m^{2}}{2k}
\\ \label{2.47'}
&&+\frac{1}{4k^{2}} \det\left(\begin{array}{cc}
A_{0}+A\cos^{2}\theta & A\sin\theta\cos\theta \\ A
\sin\theta\cos\theta &A_{0}+A\sin^{2}\theta
\end{array}\right).
\end{eqnarray}
It should be noted that, although (\ref{2.47'}) is also obtained from (\ref{2.26}) by employing (\ref{2.45'}), such a procedure includes internal inconsistency.
Through detailed considerations we can prove that, when the approximation in the extreme-relativistic case is consistently performed, we obtain the (approximate) eigenvalues which coincide with the usual ones\cite{MAN,FY}.
The detailed derivation is given in Appendix B.
However, a serious problem may arise with the energy eigenvectors.
This will be investigated in Sec.IV.

\subsection{Further remarks on derivation of the usual relations}\label{remarks}
In this subsection we give additional remarks on the implication of
the statement that the usual quantum mechanical formulas are obtained
by performing the replacement (\ref{2.25a}).

First we note the flavor neutrino field $\nu _{\rho}(x) (\rho=e, \mu,
\cdots)$ is expanded as
\begin{eqnarray}
\nu _{\rho}(x)=\frac{1}{\sqrt{V}}\sum _{\vec k,r}\lf\{\alpha _{\rho}
(kr;x^{0})u_{\rho}(kr)+\beta^{\dagger} _{\rho}(-kr;x^{0})v_{\rho}
(-kr)\ri\}e^{-i\vec k\vec x},
\end{eqnarray}
where $u_{\rho}(kr)$ and $v_{\rho}(kr)$ satisfy
\begin{eqnarray}
(i\not k+\mu _{\rho})u_{\rho}(kr)=0\,,\quad  (-i\not k+\mu _{\rho})v_{\rho}(kr)=0
\end{eqnarray}
with an arbitrary $\mu _{\rho}$\cite{FHY1,BHV};
$\beta _{\rho}(-kr;x^{0})$ and $v_{\rho}(-kr)$ are defined in
the same way as (\ref{2.13b}).
Here we are writing expressions and relations for the case of
arbitrary flavor number.
The relation of
\begin{eqnarray}
{\mib\alpha}_{F}(kr;x^{0})=
\left(\begin{array}{c}
\alpha _{e}(kr;x^{0})\\\alpha _{\mu}(kr;x^{0})\\\vdots
\end{array}\right)\,,\quad
{\mib\beta}_{F}(-kr;x^{0})=
\left(\begin{array}{c}
\beta _{e}(-kr;x^{0})\\\beta _{\mu}(-kr;x^{0})\\\vdots
\end{array}\right)
\end{eqnarray}
with ${\mib\alpha} _{M}(kr;x^{0})$ and ${\mib\beta} _{M}(-kr;x^{0})$ is given by
\begin{eqnarray}\label{2.32b}
\left(\begin{array}{c}
{\mib\alpha} _{F}(kr;x^{0})\\{\mib\beta}^{\dagger} _{F}(-kr;x^{0})
\end{array}\right)
={\cal K}(k)
\left(\begin{array}{c}
{\mib\alpha} _{M}(kr;x^{0})\\{\mib\beta}^{\dagger} _{M}(-kr;x^{0})
\end{array}\right),
\end{eqnarray}
where the matrix \cite{FHY1,FHY2} is expressed as
\begin{eqnarray}
{\cal K}(k)&=&
\left(\begin{array}{cc}
P(k)&i\Lambda(k)\\i\Lambda(k)&P(k)
\end{array}\right)
\equiv\left(\begin{array}{cc}
{\cal K}(k)_{\rho j}&{\cal K}(k)_{\rho \bar j}\\{\cal K}(k)_{\bar\rho j}&{\cal K}(k)_{\bar\rho \bar j}
\end{array}\right),\nonumber\\
& &P(k)=[P(k)_{\rho j}]=[z^{1/2}_{\rho j}\rho_{\rho j}(k)]\,,\quad
\Lambda(k)=[z^{1/2}_{\rho j}\lambda _{\rho j}(k)].
\end{eqnarray}
(As to the definition of $\rho _{\rho j}$ and $\lambda _{\rho j}$, see Appendix A.)

In {\it the extreme-relativistic case} (\ref{2.23}), we employ
$\omega _{j}=k+m_{j}^{2}/(2k), c_{j}=1-m_{j}^{2}/(8k^{2}), s_{j}=m_{j}/(2k)$;
on the contrary, it is not appropriate for us to employ
\begin{eqnarray}\label{2.33}
\rho _{\rho j}=1-\frac{(\mu _{\rho}-m_{j})^{2}}{8k^{2}}
\quad,\quad \lambda _{\rho j}=\frac{\mu _{\rho}-m_{j}}{2k},
\end{eqnarray}
since $\mu _{\rho}$ is arbitrary.
Notwithstanding this, the direct way for deriving the usually employed
quantum-mechanical relations is, as already seen partially from the
consideration in the preceding subsection B, to take (\ref{2.25a}), i.e.
\begin{eqnarray}\label{2.34a}
c_{j}\longrightarrow 1\quad,\quad s_{j}\longrightarrow 0\qquad  \forall j
\end{eqnarray}
in $H^{matt}_{(M)}(kr)$ and also
\begin{eqnarray}\label{2.34b}
\rho _{\rho j}\longrightarrow 1\quad,\quad \lambda _{\rho j}\longrightarrow
0 \qquad \forall \rho,j; \quad \mbox{ thus }\quad 
{\cal K}(k)\longrightarrow I_{2}\otimes Z^{1/2}.
\end{eqnarray}
In {\it this super-relativistic limit}, the neutrino- and the
antineutrino-operators decouple as seen from (\ref{2.17c}) and (\ref{2.17d});
${\mib\alpha} _{M}(k\uparrow;x^{0})$ and ${\mib\beta} _{M}(-k\downarrow;x^{0})$
behave like free operators (as far as $L_{int}(x)$ in (\ref{2.4}) is neglected).
We obtain the Hamiltonian for ${\mib\alpha}_{M}(k\downarrow;x^{0})$ expressed by
\begin{eqnarray}
{\cal H}^{matt}_{(M)}(k\downarrow;\alpha_{M})&=&{\mib\alpha}^{\dagger}_{M}
(k\downarrow;x^{0}){\cal E}^{matt}_{(M)}(k\downarrow){\mib\alpha}_{M}(k\downarrow;x^{0}),\\
{\cal E}^{matt}_{(M)}(k\downarrow)&=&\left(\begin{array}{cc}
k+\frac{m^{2}_{1}}{2k}+\kappa _{11}&\kappa _{12}\\\kappa _{12}&k+\frac{m^{2}_{2}}{2k}+\kappa _{22}
\end{array}\right).\label{2.35b}
\end{eqnarray}
This Hamiltonian is equal to
\begin{eqnarray}
{\cal H}^{matt}_{(F}(k\downarrow;\alpha _{F})&=&{\mib\alpha}^{\dagger}_{F}
(k\downarrow;x^{0}){\cal E}^{matt}_{(F)}(k\downarrow){\mib\alpha}_{F}(k\downarrow;x^{0}),\\
{\cal E}^{matt}_{(F)}(k\downarrow)&=&Z^{1/2}{\cal E}^{matt}_{(M)}(k\downarrow)Z^{1/2\dagger}\\
&=&k+\frac{\bar m^{2}+A_{0}}{2k}+\frac{1}{4k}
\left(\begin{array}{cc}
2A-\Delta m^{2}\cos(2\theta)&\Delta m^{2}\sin(2\theta)\\
\Delta m^{2}\sin(2\theta)&\Delta m^{2}\cos(2\theta)
\end{array}\right);\label{2.36b}
\end{eqnarray}
here $\bar m^{2}=\frac{m^{2}_{1}+m^{2}_{2}}{2}, \Delta m^{2}=m^{2}_{2}-m^{2}_{1}$.
Since left-handed neutrino are produced through the weak interaction in, e.g.,
center region of the sun, Eq.(\ref{2.36b}) coincide with Hamiltonian
employed in the usual quantum mechanical consideration (e.g.\cite{FY}).

The eigenvalues of ${\cal E}^{matt}_{(F)}(k\downarrow)$ are expressed as
\begin{eqnarray}\label{2.37a}
\tilde E_{j}(k\downarrow)=k+\frac{\tilde m^{2}_{j}(\downarrow)}{2k}
\end{eqnarray}
with the effective masses $\tilde m^{2}_{j}(\downarrow)$ given by
\begin{eqnarray}\label{2.37b}
\tilde m^{2}_{1(2)}(\downarrow)=\bar m^{2}+A_{0}+\frac{1}{2}
\left\{A-(+)\sqrt{(A-\Delta m^{2}\cos(2\theta))^{2}+(\Delta m^{2}\sin(2\theta))^{2}}\right\}
\end{eqnarray}
We define\cite{FY} the mixing angle $\tilde\theta$ and the operators
$\tilde\alpha _{j}(k\downarrow;x^{0}), j=1,2$, as
\begin{eqnarray}
\tilde Z^{1/2\dagger}{\cal E}^{matt}_{(F)}(k\downarrow)\tilde Z^{1/2}&=&
\left(\begin{array}{cc}
\tilde E_{1}(k\downarrow)&0\\0&\tilde E_{2}(k\downarrow)
\end{array}\right),\\
{\mib\alpha} _{F}(k\downarrow;x^{0})&=&\tilde Z^{1/2}
\left(\begin{array}{c}
\tilde \alpha _{1}(k\downarrow;x^{0})\\\tilde \alpha _{2}(k\downarrow;x^{0})
\end{array}\right)
\equiv \tilde Z^{1/2}\tilde{\mib\alpha} _{M}(k\downarrow;x^{0})\label{2.38b}
\end{eqnarray}
with
\begin{eqnarray}\label{2.38c}
\tilde Z^{1/2}&=&
\left(\begin{array}{cc}
\cos\tilde\theta&\sin\tilde\theta\\ -\sin\tilde\theta&\cos\tilde\theta
\end{array}\right),\\
\tilde \alpha _{j}(kj;x^{0})&=&\tilde \alpha _{j}(k\downarrow;0)
\exp(-i\tilde E_{j}(k\downarrow;;x^{0}).\end{eqnarray}
We obtain
\begin{eqnarray}\label{2.39a}
\tan\tilde\theta=\frac{\Delta m^{2}\sin(2\theta)}{-A+\Delta m^{2}
\cos(2\theta)+\sqrt{(A-\Delta m^{2}\cos(2\theta))^{2}+(\Delta m^{2}\sin(2\theta))^{2}}},\\
\left(\begin{array}{c}
\sin(2\tilde\theta)\\ \cos(2\tilde\theta)
\end{array}\right)
=\frac{1}{\sqrt{(A-\Delta m^{2}\cos(2\theta))^{2}+(\Delta m^{2}\sin(2\theta))^{2}}}
\left(\begin{array}{c}
\Delta m^{2}\sin(2\theta)\\ \Delta m^{2}\cos(2\theta)-A
\end{array}\right)\label{2.39b}
\end{eqnarray}
Eq.(\ref{2.37b}) is known to lead to a level crossing phenomenon of the
two-level problem in quantum mechanics\cite{MSW,BETHE,LZ,HP}.

From (\ref{2.38b}) or by integrating Heisenberg equation (without $L_{int}(x)$)
\begin{eqnarray}\label{2.40a}
i\frac{d}{dx^{0}}{\mib\alpha}_{F}(k\downarrow;x^{0})&=&[{\mib\alpha}_{F}
(k\downarrow;x^{0}),{\cal H}^{matt}_{(F)}(k\downarrow;{\mib\alpha}_{F})]\nonumber\\
&=&\tilde Z^{1/2}
\left(\begin{array}{cc}
\tilde E_{1}(k\downarrow)&0\\0&\tilde E_{2}(k\downarrow)
\end{array}\right)
\tilde Z^{1/2\dagger}{\mib\alpha}_{F}(k\downarrow;x^{0}).
\end{eqnarray}
we obtain\cite{FY}
\begin{eqnarray}\label{2.40b}
{\alpha}_{\rho}(k\downarrow;x^{0})=\sum _{\sigma,j}\tilde z^{1/2}_{\rho j}
\exp(-i\tilde E_{j}(k\downarrow)(x^{0}-x^{0}_{I}))\cdot\tilde
z^{1/2\ast}_{\sigma j}\alpha_{\sigma}(k\downarrow;x^{0}_{I}).
\end{eqnarray}
When we define the vacuum $|\tilde 0\rangle$ as
\begin{eqnarray}
\tilde\alpha _{\rho}(k\downarrow;0)|\tilde 0\rangle =0\,\,\, \mbox{ for } \forall\vec k,\rho,
\end{eqnarray}
and the state
\begin{eqnarray}
\Psi _{\rho}(k\downarrow;x^{0})=\langle\tilde 0|\alpha _{\rho}(k\downarrow;x^{0}),
\end{eqnarray}
we obtain Schr\"odinger equation corresponding to (\ref{2.40a}) and its integration
\begin{eqnarray}
\Psi _{\rho}(k\downarrow;x^{0})=\sum _{\sigma,j}\tilde z^{1/2}_{\rho j}
\exp\left[-i\tilde E_{j}(k\downarrow)(x^{0}-x^{0}_{I})\right]\cdot\tilde z^{1/2\ast}_{\sigma j}
\Psi _{\sigma}(k\downarrow;x^{0}_{I}).
\end{eqnarray}
Then, the transition amplitudes for $\nu _{\sigma}\rightarrow \nu _{\rho}$ in matter are
\begin{eqnarray}\label{2.42a}
A^{matt}_{\rho\sigma}(k\downarrow;x^{0}-x^{0}_{I})=\sum _{j}\tilde z^{1/2}_{\rho j}
\exp\left[-i\tilde E_{j}(k\downarrow)(x^{0}-x^{0}_{I})\right]\cdot\tilde z^{1/2\ast}_{\sigma j}
\end{eqnarray}
and their absolute squares lead to the transition probabilities
\begin{eqnarray}\label{2.42b}
P^{matt}(\nu _{e}\rightarrow\nu _{\mu})=\frac{1}{2}\sin^{2}(2\tilde\theta)
[1-\cos(\Delta E^{matt}L)]=P^{matt}(\nu _{\mu}\rightarrow\nu _{e}),\\
P^{matt}(\nu _{e}\rightarrow\nu _{e})=P^{matt}(\nu _{\mu}\rightarrow\nu _{\mu})
=1-P^{matt}(\nu _{e}\rightarrow\nu _{\mu});\label{2.42c}
\end{eqnarray}
here $L=(x^{0}-x^{0}_{L})/c$, the approximate distance that neutrino passes
in matter;$\Delta E^{matt}=\tilde m_{2}(\downarrow)^{2}-\tilde m_{1}(\downarrow)^{2}$.

In the same procedure, we obtain Hamiltonian for ${\mib\beta}_{F}(-k\uparrow;x^{0})$ given by
\begin{eqnarray}
{\cal H}^{matt}_{(F)}(k\uparrow;{\mib\beta} _{F})&=&
{\mib\beta}^{\dagger}_{F}(-k\uparrow;x^{0}){\cal E}^{matt}_{(F)}
(k\uparrow){\mib\beta}_{F}(-k\uparrow;x^{0}),\\
{\cal E}^{matt}_{(F)}(k\uparrow)&=&Z^{1/2\ast}{\cal E}^{matt}_{(M)}(k\uparrow)(Z^{1/2})^{T},
\end{eqnarray}
where ${\cal E}^{matt}_{(M)}(k\uparrow)$ is expressed as
RHS of (\ref{2.35b}) with $-\kappa _{jl}$ substituted for $\kappa _{jl}$;
thus ${\cal E}^{matt}_{(F)}(k\uparrow)$ is given by RHS of (\ref{2.36b})
with $-A_{0}$ and $-A$ instead of $A_{0}$ and $A$.

As pointed out in the preceding subsection B and concerning (\ref{2.33}),
the substitutions (\ref{2.34a}) and (\ref{2.34b}) are problematic approximations.
Section IV will be devoted to examining this problem.

\section{Poles of flavor-neutrino propagator}\label{Poles}
We give a remark on the poles of the flavor-neutrino propagator
and their relation with the energy eigenvalues obtained in
Sec.II.B. Those poles correspond to the physical masses of
neutrinos\cite{KOW}. It has been shown in Ref.\cite{FHY1}
concretely in the two- and three-flavor neutrinos in vacuum that
the diagonalization of the pole part in the propagator matrix is
equivalent to the diagonalization of the mass matrix in Lagrangian
in so far as the propagator matrix is evaluated by taking into
account fully the repetition of the bilinear (mass-type)
interaction and dropping other interactions. This is, as noted in
Ref.\cite{FHY1}, independent of the ways separating the bilinear
interaction part from the free Lagrangian.

Now we consider the neutrino propagation in matter, where the Lagrangian (\ref{2.4}) is
rewritten as
\begin{eqnarray}\label{3.1a}
L^{matt}_{eff}(x)&=&-
\left(\begin{array}{cc}
\bar\nu _{e}&\bar\nu _{\mu}
\end{array}\right)
\lf[\not\partial +
\left(\begin{array}{cc}
\hat m_{ee}&0\\0&\hat m_{\mu\mu}
\end{array}\right)\ri]
\left(\begin{array}{c}
\nu _{e}\\ \nu _{\mu}
\end{array}\right)
+L_{I}(x)+L_{int}(x),\\
L_{I}(x)&=&
-\left(\begin{array}{cc}
\bar\nu _{e}&\bar\nu _{\mu}
\end{array}\right)
\left(\begin{array}{cc}
\Delta _{ee}-i(b_{0}+b_{1})\not u P_{L}&m_{e\mu}\\m_{e\mu}&\Delta _{\mu\mu}-ib_{0}\not u P_{L})
\end{array}\right)
\left(\begin{array}{c}
\nu _{e}\\ \nu _{\mu}
\end{array}\right)
\end{eqnarray}
with $\Delta _{\rho\rho}=m_{\rho\rho}-\hat m_{\rho\rho}$.

The physical masses of neutrinos are obtained as poles of the propagator
$S'_{\sigma\rho}(k)$ of the unrenormalized Heisenberg fields $\nu _{\rho}(x)$'s;
\begin{eqnarray}
S'_{\sigma\rho}(x-y)=\langle 0|T(\nu _{\sigma}(x)\bar\nu _{\rho}(y))|0\rangle =
\frac{1}{(2\pi)^{4}}\int d^{4}k e^{ik(x-y)}S'_{\sigma\rho}(k).
\end{eqnarray}
The Fourier transform $S'_{\sigma\rho}(k)$ satisfies
\begin{eqnarray}\label{3.3}
S'_{\sigma\rho}(k)=\delta _{\sigma\rho}S_{\rho}(k)+\sum _{\lambda}
S'_{\sigma\lambda}(k)\Pi _{\lambda\rho}(k)S_{\rho}(k);
\end{eqnarray}
here, $\Pi _{\lambda\rho}(k)$ and $S_{\rho}(k)$ are the proper self-energy
part and the Fourier transform of the free propagator when the first part in
RHS of (\ref{3.1a}) is regarded as the free Lagrangian.
Then $S_{\rho}(k)$ is expressed as
\begin{eqnarray}
S_{\rho}(k)=\frac{-\not k -i\hat m_{\rho\rho}}{k^{2}+\hat m^{2}_{\rho\rho}-i\epsilon}.
\end{eqnarray}
We neglect $L_{int}(x)$ which is taken not to include any bilinear terms of the
flavor neutrino fields, and the repetition of $L_{I}(x)$ is fully taken into account;
then we obtain (by assuming the x-independence of $b_{0}$ and $b_{1}$)
\begin{eqnarray}
[\Pi _{\sigma\rho}(k)]=-i
\left(\begin{array}{cc}
\Delta _{ee}-i(b_{0}+b_{1})\not u P_{L}&m_{e\mu}\\m_{e\mu}&\Delta _{\mu\mu}-ib_{0}\not u P_{L}
\end{array}\right)
\end{eqnarray}

When we define the matrix $[f_{\sigma\rho}(k)]$ to be
\begin{eqnarray}
S' _{\sigma\rho}(k)=[f(k)^{-1}]_{\sigma\rho},
\end{eqnarray}
we obtain from (\ref{3.3})
\begin{eqnarray}
f_{\sigma\rho}(k)=\delta _{\sigma\rho}S_{\sigma}(k)^{-1}-\Pi _{\sigma\rho}(k),
\end{eqnarray}
the concrete form of which is
\begin{eqnarray}
[f_{\sigma\rho}(k)]=
\left(\begin{array}{cc}
-\not k+i\{m_{ee}-i(b_{0}+b_{1})\not u P_{L}\}&im_{e\mu}\\im_{e\mu}&-\not
k+i\{m _{\mu\mu}-ib_{0}\not u P_{L}\}
\end{array}\right),
\end{eqnarray}
or
\begin{eqnarray}\label{3.7c}
\frac{1}{i}[f_{\sigma\rho}(k)]=
\left(\begin{array}{cccc}
m_{ee}&-(\vec\sigma\vec k+k_{0})+N&m_{e\mu}&0\\
\vec\sigma\vec k-k_{0}&m_{ee}&0&m_{e\mu}\\
m_{e\mu}&0&m_{\mu\mu}&-(\vec\sigma\vec k+k_{0})+N'\\
0&m_{e\mu}&\vec\sigma\vec k-k_{0}&m_{\mu\mu}
\end{array}\right)
\end{eqnarray}
where $N$ and $N'$ are defined in (\ref{2.5}).
We see that, as expected, $[f_{\sigma\rho}(k)]$ is free from the
arbitrariness of separating $L_{I}$ in $L^{matt}_{eff}$.

In the same way, from Lagrangian (\ref{2.10a}) expressed in terms of
the mass-basis operators, we can define $S'_{jl}(k)$ which is the Fourier transform of
\begin{eqnarray}
S'_{jl}(x-y)=\langle 0|T(\nu _{j}(x)\bar\nu _{l}(y))|0\rangle
\end{eqnarray}
We obtain
\begin{eqnarray}\label{3.8b}
\frac{1}{i}[f^{(m)}_{jl}(k)]&\equiv&\frac{1}{i}[(S'(k)^{-1})_{jl}]\nonumber\\
&=&
\left(\begin{array}{cccc}
m_{1}&-(\vec\sigma\vec k+k_{0})+B&0&G\\
\vec\sigma\vec k-k_{0}&m_{1}&0&0\\
0&G&m_{2}&-(\vec\sigma\vec k+k_{0})+B'\\
0&0&\vec\sigma\vec k-k_{0}&m_{2}
\end{array}\right);
\end{eqnarray}
here $B$, $B'$ and $G$ are defined in (\ref{2.16'}).
We can confirm concretely the relation between (\ref{3.7c}) and (\ref{3.8b})
\begin{eqnarray}
[f^{(m)}_{jl}(k)]=Z^{1/2\dagger}[f_{\sigma\rho}(k)]Z^{1/2}\quad , \quad Z^{1/2}=
\left(\begin{array}{cc}
\cos\theta&\sin\theta\\-\sin\theta&\cos\theta
\end{array}\right),
\end{eqnarray}
expected from $\nu_{\rho}(x)=\sum_{j}z^{1/2}_{\rho j}\nu _{j}(x)$.

Next we examine the poles of the propagator $S'(k)$, which are given
by $\det[f_{\sigma\rho}(k)]=\det[f^{(m)}_{jl}(k)]=0$.
It is useful for us to note that, for a matrix $\left(
\begin{array}{cc}a&c\\d&b\end{array}\right)$ with square sub-matrices
$a, b, c$ and $d$, we have
\begin{eqnarray}
\det\left(\begin{array}{cc}
a&c\\d&b
\end{array}\right)=
\left|\left(\begin{array}{cc}
a&0\\d&I
\end{array}\right)
\left(\begin{array}{cc}
I&a^{-1}c\\0&b-da^{-1}c
\end{array}\right)\right|,
\end{eqnarray}
which is equal to $|ab-dc|$ if $[a^{-1}, d]=0$.
\footnote{
Note that
\begin{eqnarray}
det\left[\frac{1}{i}f_{jl}^{m}(k)\right]=det\left(\begin{array}{cccc}
m_{1}&0&-(\vec\sigma\vec k+k_{0})+B&G\\
0&m_{2}&G&-(\vec\sigma\vec k+k_{0})+B'\\
\vec\sigma\vec k-k_{0}&0&m_{1}&0\\
0&\vec\sigma\vec k-k_{0}&0&m_{2}\\
\end{array}\right).
\end{eqnarray}
}
We obtain
\begin{eqnarray}\label{3.10a}
\det[\frac{1}{i}f^{(m)}_{jl}(k)]=&&K^{2}_{1}K^{2}_{2}
+2K_{1}K_{2}\{(K_{1}\kappa _{22}+K_{2}\kappa _{11})(\vec k\vec u-k_{0}u_{0})
+(\det\kappa)[2(\vec k\vec u-k_{0}u_{0})^{2}-k^{2}_{0}+\vec k^{2}]\}\nonumber\\
&&+(k^{2}_{0}-\vec k^{2})\{(K_{1}\kappa _{22}+K_{2}\kappa _{11})^{2}
+2(K_{1}\kappa _{22}+K_{2}\kappa _{11})(\det\kappa)(\vec k\vec u-k_{0}u_{0})
+(\det\kappa)^{2}(k^{2}_{0}-\vec k^{2})\},
\end{eqnarray}
where $K_{j}=k^{2}_{0}-(\vec k^{2}+m^{2}_{j})=k^{2}_{0}-\omega^{2}_{j}$.
RHS of (\ref{3.10a}) is divided into two parts, one of which vanishes in the static medium:
\begin{eqnarray}
(\ref{3.10a})
&=&\left[K^{2}_{1}K^{2}_{2}+2K_{1}K_{2}\{-k_{0}(K_{1}\kappa _{22}+
K_{2}\kappa _{11})+(\det\kappa)(k^{2}_{0}+\vec k^{2})\}\right.\nonumber\\
& &\left.+(k^{2}_{0}-\vec k^{2})\lf\{(K_{1}\kappa _{22}+
K_{2}\kappa _{11})^{2}-2k_{0}(\det\kappa)(K_{1}\kappa _{22}+
K_{2}\kappa _{11})+(\det\kappa)^{2}(k^{2}_{0}-\vec k^{2})\ri\}\right]\nonumber\\
&&+\left[2K_{1}K_{2}\lf\{(K_{1}\kappa _{22}+K_{2}\kappa _{11})
(\vec k\vec u-k_{0}u_{0}+k_{0})+2(\det\kappa)[(\vec k\vec u-k_{0}u_{0})^{2}-k^{2}_{0}]
\right.\ri\}\nonumber\\
&&\left.+2(k^{2}_{0}-\vec k^{2})(K_{1}\kappa _{22}+K_{2}\kappa _{11})
(\det\kappa)(\vec k\vec u-k_{0}u_{0}+k_{0})\right].
\end{eqnarray}
Thus we have
\begin{eqnarray}
Det(k)_{static}&\equiv&\det[\frac{1}{i}f^{(m)}_{jl}(k)]_{static}=
\det[\frac{1}{i}f_{\sigma\rho}(k)]_{static}
\nonumber\\
&=&\lf[(k^{2}_{0}-\omega^{2}_{1})(k^{2}_{0}-\omega^{2}_{2})-(k_{0}-
|\vec k|)\lf\{(k^{2}_{0}-\omega^{2}_{1})\kappa _{22}+(k^{2}_{0}-
\omega^{2}_{2})\kappa _{11}-(k_{0}-|\vec k|)(\det\kappa)\ri\}\ri]
\nonumber\\
&&\times\lf[(k^{2}_{0}-\omega^{2}_{1})(k^{2}_{0}-\omega^{2}_{2})-
(k_{0}+|\vec k|)\lf\{(k^{2}_{0}-\omega^{2}_{1})\kappa _{22}+
(k^{2}_{0}-\omega^{2}_{2})\kappa _{11}-(k_{0}+|\vec k|)(\det\kappa)\ri\}\ri].
\label{3.10c}
\end{eqnarray}

We see that the two parts composing $Det(|\vec k|,k_{0}=\lambda)_{static}$
are equal to (\ref{2.19a}) and (\ref{2.19b}), respectively, i.e.
\begin{eqnarray}
Det(|\vec k|,k_{0}=\lambda)_{static}=\det[(H^{matt}_{(M)}(k\uparrow)-
\lambda)(H^{matt}_{(M)}(k\downarrow)-\lambda)]
\end{eqnarray}
This is a natural equality in the present approximation of neglecting
$L_{int}(x)$ contribution, since ${\cal H}^{matt}_{(M)}(x^{0})$ in
the static medium is separated into helicity-up and -down parts like (\ref{2.17a}).

\section{Energy-eigenvectors in matter}\label{eigenvectors}
We investigate the eigenvectors of $H^{matt}_{(M)}(k\downarrow)$
and $H^{matt}_{(F)}(k\downarrow)$ corresponding to the eigenvalues
 in the extreme-relativistic case obtained in Appendix B.
For convenience, we write these eigenvalues as
\begin{eqnarray}
E_{a}(k\downarrow)&\equiv&k+\frac{\tilde m_{(+)a}(\downarrow)^{2}}{2k},\\
E_{a+2}(k\downarrow)&\equiv&-(k+\frac{\tilde m_{(-)a}(\downarrow)^{2}}{2k}), \,\,\,a=1,2.,
\end{eqnarray}

Here we give a concrete form of $H^{matt}_{(F)}(k\downarrow)$in the flavor basis;
this is obtained by utilizing (\ref{2.17b}) and
\begin{eqnarray}
H^{matt}_{(F)}(k\downarrow)={\cal K}(k)H^{matt}_{(M)}(k){\cal K}(k\downarrow)^{\dagger},
\end{eqnarray}
derived from (\ref{2.17b}) and (\ref{2.32b});
\begin{eqnarray}
&&H^{matt}_{(F)}(k\downarrow)=\\ \nonumber
&&\left(\begin{array}{cccc} \omega _{e}+\frac{\mu _{e}(m_{ee}-\mu
_{e})}{\omega _{e}}+Nc_{e}c_{e} &m_{e\mu}\sin\frac{\chi _{e}+\chi
_{\mu}}{2} &i\frac{k(m_{ee}-\mu _{e})}{\omega _{e}}-ic_{e}s_{e}N
&im_{e\mu}\cos\frac{\chi _{e}+\chi _{\mu}}{2}\\
m_{e\mu}\sin\frac{\chi _{e}+\chi _{\mu}}{2}&\omega _{\mu}+
\frac{\mu _{\mu}(m_{\mu\mu}-\mu _{\mu})}{\omega _{\mu}}+
N'c_{\mu}c_{\mu}&im_{e\mu}\cos\frac{\chi _{e}+\chi _{\mu}}{2}&
i\frac{k(m_{\mu\mu}-\mu _{\mu})}{\omega _{\mu}}-ic_{\mu}s_{\mu}N'\\
-i\frac{k(m_{ee}-\mu _{e})}{\omega _{e}}+iNs_{e}c_{e}
&-im_{e\mu}\cos\frac{\chi _{e}+\chi _{\mu}}{2}&-\omega _{e}-
\frac{\mu _{e}(m_{ee}-\mu _{e})}{\omega _{e}}+
Ns_{e}s_{e}&-m_{e\mu}\sin\frac{\chi _{e}+\chi _{\mu}}{2}\\
-im_{e\mu}\cos\frac{\chi _{e}+\chi _{\mu}}{2}&-
i\frac{k(m_{\mu\mu}-\mu _{\mu})}{\omega _{\mu}}+
iN's_{\mu}c_{\mu}&-m_{e\mu}\sin\frac{\chi _{e}+\chi _{\mu}}{2}&-
\omega _{\mu}-\frac{\mu _{\mu}(m_{\mu\mu}-\mu _{\mu})}{\omega _{\mu}}+N's_{\mu}s_{\mu}
\end{array}\right),
\end{eqnarray}
where $\omega _{\rho}=\sqrt{k^{2}+\mu^{2}_{\rho}}$,
$s_{\rho}=\sin(\chi _{\rho}/2)$, $c_{\rho}=\cos(\chi _{\rho}/2)$,
$\cot\chi _{\rho}=k/\mu _{\rho}$,
 $N=b_{0}+b_{1}$, $N'=b_{0}$.
Thus we see that the eigenvectors of $H^{matt}_{(F)}(k\downarrow)$
do depend generally on a set of arbitrary mass
parameters $\{\mu _{e}, \mu _{\mu}\}$.
We will examine this situation.
\subsection{general relations}\label{grelations}
Let us define a set of the eigenvectors as
\begin{eqnarray}\label{5.3a}
H^{matt}_{(M)}(k\downarrow){\mib y}^{(a)}_{M}&=
&E_{a}(k\downarrow){\mib y}^{(a)}_{M},\\
H^{matt}_{(F)}(k\downarrow){\mib y}^{(a)}_{F}&=
&E_{a}(k\downarrow){\mib y}^{(a)}_{F}\quad, \quad a=1,2,3,4;
\end{eqnarray}
here, we take these eigenvectors so as to satisfy the ortho-normality condition
\begin{eqnarray}
{\mib y}^{(a)\dagger}_{M}{\mib y}^{(b)}_{M}=\delta _{ab}\,,\quad
{\mib y}^{(a)\dagger}_{F}{\mib y}^{(b)}_{F}=\delta _{ab},
\end{eqnarray}
and to have the relation
\begin{eqnarray}
{\mib y}^{(a)}_{F}={\cal K}(k){\mib y}^{(a)}_{M}.
\end{eqnarray}
By defining the unitary matrices
\begin{eqnarray}\label{5.5a}
T_{(M)}(k)&\equiv&
\left(\begin{array}{cccc}
{\mib y}^{(1)}_{M}&{\mib y}^{(2)}_{M}&{\mib y}^{(3)}_{M}&{\mib y}^{(4)}_{M}
\end{array}\right),\\
T_{(F)}(k)&\equiv&
\left(\begin{array}{cccc}
{\mib y}^{(1)}_{F}&{\mib y}^{(2)}_{F}&{\mib y}^{(3)}_{F}&{\mib y}^{(4)}_{F}
\end{array}\right)
={\cal K}(k)T_{(M)}(k),
\end{eqnarray}
we have
\begin{eqnarray}
T^{\dagger}_{(M)}(k)H^{matt}_{(M)}(k\downarrow)T_{(M)}(k)=
\left(\begin{array}{cccc}
E_{1}(k\downarrow)&0&0&0\\
0&E_{2}(k\downarrow)&0&0\\
0&0&E_{3}(k\downarrow)&0\\
0&0&0&E_{4}(k\downarrow)
\end{array}\right)
\equiv E^{matt}_{diag}(k\downarrow);
\end{eqnarray}
thus, from (\ref{2.17b}),
\begin{eqnarray}\label{5.6b}
{\cal H}^{matt}_{(M)}(k\downarrow;x^{0})=
\left(\begin{array}{cc}
{\mib\alpha}^{\dagger}_{M}(k\downarrow;x^{0})&{\mib\beta}_{M}(-k\downarrow;x^{0})
\end{array}\right)
T_{(M)}(k)E^{matt}_{diag}(k\downarrow)T_{(M)}^{\dagger}(k)
\left(\begin{array}{c}
{\mib\alpha}_{M}(k\downarrow;x^{0})\\{\mib\beta}^{\dagger}_{M}(-k\downarrow;x^{0})
\end{array}\right)
\end{eqnarray}
Here, for convenience, we redefine $H^{matt}_{(M)}(k\downarrow;x^{0})$
by adding an appropriate c-number so that the normal ordering symbol
$:\,\,  :$ is dropped.
The above expression is rewritten as
\begin{eqnarray}\label{5.7}
\mbox{RHS of (\ref{5.6b})}=
\left(\begin{array}{cc}
{\mib\alpha}^{\dagger}_{F}(k\downarrow;x^{0})&{\mib\beta}_{F}(-k\downarrow;x^{0})
\end{array}\right)
T_{(F)}(k)E^{matt}_{diag}(k\downarrow)T_{(F)}^{\dagger}(k)
\left(\begin{array}{c}
{\mib\alpha}_{F}(k\downarrow;x^{0})\\{\mib\beta}^{\dagger}_{F}(-k\downarrow;x^{0})
\end{array}\right)
\equiv {\cal H}^{matt}_{(F)}(k\downarrow;x^{0}),
\end{eqnarray}
which is the Hamiltonian in the flavor basis.

Corresponding to (\ref{2.38b}), we define renewdly the operators (with $a=1,2$):
\begin{eqnarray}
\tilde{\mib\alpha}_{a}(k\downarrow;x^{0})&\equiv&{\mib y}^{(a)}_{F}(k)^{\dagger}
\left(\begin{array}{c}
{\mib\alpha}_{F}(k\downarrow;x^{0})\\{\mib\beta}^{\dagger}_{F}(-k\downarrow;x^{0})
\end{array}\right),\\
\tilde{\mib\beta}^{\dagger}_{a}(-k\downarrow;x^{0})&\equiv&{\mib y}^{(a+2)}_{F}(k)^{\dagger}
\left(\begin{array}{c}
{\mib\alpha}_{F}(k\downarrow;x^{0})\\{\mib\beta}^{\dagger}_{F}(-k\downarrow;x^{0})
\end{array}\right);
\end{eqnarray}
then we write them as
\begin{eqnarray}\label{5.8c}
\left(\begin{array}{c}
\tilde{\mib\alpha}_{M}(k\downarrow;x^{0})\\\tilde{\mib\beta}^{\dagger}_{M}(-k\downarrow;x^{0})
\end{array}\right)
=T^{\dagger}_{(F)}
\left(\begin{array}{c}
{\mib\alpha}_{F}(k\downarrow;x^{0})\\{\mib\beta}^{\dagger}_{F}(-k\downarrow;x^{0})
\end{array}\right)
=T^{\dagger}_{(M)}
\left(\begin{array}{c}
{\mib\alpha}_{M}(k\downarrow;x^{0})\\{\mib\beta}^{\dagger}_{M}(-k\downarrow;x^{0})
\end{array}\right),
\end{eqnarray}
and Hamiltonian (\ref{5.6b}) (or (\ref{5.7})) is expressed in
terms of $\tilde{\mib\alpha}_{M}(k\downarrow;x^{0})$ and
$\tilde{\mib\beta}_{M}(-k\downarrow;x^{0})$ as
\begin{eqnarray}\label{5.8d}
{\cal H}^{matt}_{(M)}(k\downarrow;x^{0})={\cal H}^{matt}_{(F)}(k\downarrow;x^{0})=
\left(\begin{array}{cc}
\tilde{\mib\alpha}^{\dagger}_{M}(k\downarrow;x^{0})&
\tilde{\mib\beta}_{M}(-k\downarrow;x^{0})
\end{array}\right)
E^{matt}_{diag}(k\downarrow)
\left(\begin{array}{c}
\tilde{\mib\alpha}_{M}(k\downarrow;x^{0})\\
\tilde{\mib\beta}^{\dagger}_{M}(-k\downarrow;x^{0})
\end{array}\right).
\end{eqnarray}
By noting that
\begin{eqnarray}
\tilde\alpha_{a}(k\downarrow;x^{0})&=&e^{-iE_{a}(k\downarrow)x^{0}}
\tilde\alpha_{a}(k\downarrow;0),
\nonumber\\
\tilde\beta_{a}(-k\downarrow;x^{0})&=&e^{iE_{a+2}(k\downarrow)x^{0}}
\tilde\beta_{a}(-k\downarrow;0)
\end{eqnarray}
we see that (\ref{5.8d}) is time-independent;
then we write it as
\begin{eqnarray}
{\cal H}^{matt}(k\downarrow)&=&
\left(\begin{array}{cc}
\tilde{\mib\alpha}^{\dagger}_{M}(k\downarrow)&\tilde{\mib\beta}_{M}(-k\downarrow)
\end{array}\right)
E^{matt}_{diag}(k\downarrow)
\left(\begin{array}{c}
\tilde{\mib\alpha}_{M}(k\downarrow)\\\tilde{\mib\beta}^{\dagger}_{M}(-k\downarrow)
\end{array}\right),
\end{eqnarray}
where
\begin{eqnarray}
\left(\begin{array}{c}
\tilde{\mib\alpha}_{M}(k\downarrow)\\\tilde{\mib\beta}^{\dagger}_{M}(-k\downarrow)
\end{array}\right)&\equiv&
\left(\begin{array}{c}
\tilde{\mib\alpha}_{M}(k\downarrow;0)\\\tilde{\mib\beta}^{\dagger}_{M}(-k\downarrow;0)
\end{array}\right).
\end{eqnarray}

Now we consider Heisenberg equation
\begin{eqnarray}
i\frac{d}{dx^{0}}
\left(\begin{array}{c}
{\mib\alpha}_{F}(k\downarrow;x^{0})\\{\mib\beta}^{\dagger}_{F}(-k\downarrow;x^{0})
\end{array}\right)
&=&[\left(\begin{array}{c}
{\mib\alpha}_{F}(k\downarrow;x^{0})\\{\mib\beta}^{\dagger}_{F}(-k\downarrow;x^{0})
\end{array}\right),\mbox{Eq.(\ref{5.7})}]=H^{matt}_{(F)}(k\downarrow)
\left(\begin{array}{c}
{\mib\alpha}_{F}(k\downarrow;x^{0})\\{\mib\beta}^{\dagger}_{F}(-k\downarrow;x^{0})
\end{array}\right),
\end{eqnarray}
where
\begin{eqnarray}
H^{matt}_{(F)}(k\downarrow)&=&T_{(F)}(k)E^{matt}_{diag}(k\downarrow)T^{\dagger}_{(F)}(k).
\end{eqnarray}
By defining the vacuum $|\tilde 0\rangle$ as
\begin{eqnarray}
\left\{\begin{array}{c}
{\tilde\alpha}_{a}(k\downarrow;x^{0})\\{\tilde\beta}_{a}(-k\downarrow;x^{0})
\end{array}\right\}|\tilde 0\rangle =0 \,\qquad \mbox{ for }\qquad\forall \vec k,a, x_{0}
\end{eqnarray}
and
\begin{eqnarray}
\Psi _{\rho}(k\downarrow;x^{0})\equiv \langle \tilde 0|\alpha _{\rho}(k\downarrow;x^{0}),
\end{eqnarray}
it may be possible for us, similarly to the explanation in
Sec.II.C, to investigate the evolution relations.
There will appears, however, a certain new aspect different
from the simple relation (\ref{2.40b}):
Even in the approximation framework of the extreme-relativistic
case employed for obtaining the energy eigenvalues of $H^{matt}_{(M)}
(k\downarrow)$, the matrix $T_{(F)}(k)$ giving a sort of evolution relation
\begin{eqnarray}\label{5.12a}
\left(\begin{array}{c}
{\mib\alpha}_{F}(k\downarrow;x^{0})\\{\mib\beta}^{\dagger}_{F}(-k\downarrow;x^{0})
\end{array}\right)
&=&T_{(F)}(k)
\left(\begin{array}{c}
\tilde{\mib\alpha}_{M}(k\downarrow;x^{0})\\\tilde{\mib\beta}^{\dagger}_{M}(-k\downarrow;x^{0})
\end{array}\right)
=W^{(F)}(k\downarrow;x^{0}-x^{0}_{I})^{\dagger}
\left(\begin{array}{c}
{\mib\alpha}_{F}(k\downarrow;x^{0}_{I})\\{\mib\beta}^{\dagger}_{F}(-k\downarrow;x^{0}_{I})
\end{array}\right),\\
W^{(F)}(k\downarrow;x^{0}-x^{0}_{I})&\equiv&T_{(F)}(k)
\Phi(k;x^{0}-x^{0}_{I})T^{\dagger}_{(F)}(k),\\
\Phi(k;x^{0})&\equiv&\exp(iE^{matt}_{diag}(k\downarrow)x^{0}),
\end{eqnarray}
not only depends on the arbitrary mass parameters $\mu _{\lambda}$
but also has {\it off-diagonal} elements connecting $\tilde{\mib\alpha}_{M}
(k\downarrow;x^{0})$ to $\tilde{\mib\beta}^{\dagger}_{M}(-k\downarrow;x^{0})$.
In the following subsection we examine such a situation concretely.
\subsection{Approximate forms of eigenvectors}\label{appeigenvectors}
By employing (\ref{2.17d}) , Eq.(\ref{5.3a})  for $a=1$ case is explicitly written as
\begin{eqnarray}\label{5.13}
\frac{1}{2k}
&&\left(\begin{array}{cc}
m^{2}_{1}-\tilde m_{(+)1}(\downarrow)^{2}+(A_{0}+A c^{2}_{\theta})c_{1}c_{1}
&A c_{\theta}s_{\theta}c_{1}c_{2}\\
A c_{\theta}s_{\theta}c_{2}c_{1}&m^{2}_{2}-\tilde m_{(+)1}(\downarrow)^{2}+
(A_{0}+A s^{2}_{\theta})c_{2}c_{2}\\
i(A_{0}+Ac^{2}_{\theta})s_{1}c_{1}&iAc_{\theta}s_{\theta}s_{1}c_{2}\\
iAc_{\theta}s_{\theta}s_{2}c_{1}&i(A_{0}+As^{2}_{\theta})s_{2}c_{2}
\end{array}\right.\nonumber\\
\nonumber\\
&&\,\,\,\,\left.\begin{array}{cc}
-i(A_{0}+A c^{2}_{\theta})c_{1}s_{1}&-i A c_{\theta}s_{\theta}c_{1}s_{2}\\
-i A c_{\theta}s_{\theta}c_{2}s_{1}&-i(A_{0}+A s^{2}_{\theta})c_{2}s_{2}\\
-4k^{2}-m^{2}_{1}-\tilde m_{(+)1}(\downarrow)^{2}+(A_{0}+A c^{2}_{\theta})
s_{1}s_{1}&A c_{\theta}s_{\theta}s_{1}s_{2}\\
Ac_{\theta}s_{\theta}s_{2}s_{1}&-4 k^{2}-m^{2}_{2}-\tilde m_{(+)1}(\downarrow)^{2}+
(A_{0}+A s^{2}_{\theta})s_{2}s_{2}
\end{array}\right)
{\mib y}^{(1)}_{M}=0.
\end{eqnarray}
In the extreme-relativistic case, we see from the third and the forth equations that
\begin{eqnarray}
y^{(1)}_{M,3}, \,\, y^{(1)}_{M,4}={\cal O}(1/k^{n})\quad, \quad n\geq 3
\end{eqnarray}
in comparison with $1$.
Then, from the first and the second equations of (\ref{5.13}) we obtain,
by dropping $O(1/k^{2})$-corrections to the masses
\begin{eqnarray}\label{5.14b}
\left(\begin{array}{cc}
m^{2}_{1}-\tilde m_{(+)1}(\downarrow)^{2}+A_{0}+Ac^{2}_{\theta}&
Ac_{\theta}s_{\theta}\\Ac_{\theta}s_{\theta}&m^{2}_{2}-\tilde m_{(+)1}
(\downarrow)^{2}+A_{0}+As^{2}_{\theta}
\end{array}\right)
\left(\begin{array}{c}
y^{(1)}_{M,1}\\y^{(1)}_{M,2}
\end{array}\right)=0.
\end{eqnarray}
The condition  for the matrix on LHS  to be consistent
with $|y^{(1)}_{M,1}|^{2}+|y^{(1)}_{M,2}|^{2}=1$ is its determinant to be equal to $0$.
This condition is seen to be
satisfied, because, due to (\ref{4.7}) , the determinant is written as
\begin{eqnarray}\label{5}
\frac{1}{4}&&\lf\{-\Delta m^{2}+A\cos(2\theta)+\sqrt{(A\cos(2\theta)-
\Delta m^{2})^{2}+(A\sin(2\theta))^{2}}\ri\}\nonumber\\
\times&&\lf\{\Delta m^{2}-A\cos(2\theta)+\sqrt{(A\cos(2\theta)-
\Delta m^{2})^{2}+(A\sin(2\theta))^{2}}\ri\}-(\frac{1}{2}A\sin(2\theta))^{2}=0.
\end{eqnarray}

As for the vector ${\mib y}^{(2)}_{M}$, we see that, due to  the condition with $(-)$
sign substituted for the sign $(+)$ in front of the square root in (\ref{5}), the solution
\begin{eqnarray}
y^{(2)}_{M,3},\,\,\, y^{(2)}_{M,4}={\cal O}(1/k^{n})\quad, \quad n\geq 3, \qquad
|y^{(2)}_{M,1}|^{2}+|y^{(2)}_{M,2}|^{2}=1
\end{eqnarray}
is obtained.

As for ${\mib y}^{(3)}_{M}$ and ${\mib y}^{(4)}_{M}$, possible solutions are seen to exist:
for $n\geq 2$
\begin{eqnarray}
y^{(3)}_{M,3}&=&{\cal O}(k^{0})\quad; \quad y^{(3)}_{M,1}, y^{(3)}_{M,2}=
{\cal O}(1/k^{n+1})\quad; \quad  y^{(3)}_{M,4}={\cal O}(1/k^{n}),
\\
y^{(4)}_{M,3}&=&{\cal O}(k^{0})\quad; \quad  y^{(4)}_{M,1}, y^{(4)}_{M,2}=
{\cal O}(1/k^{n+1})\quad; \quad  y^{(4)}_{M,3}={\cal O}(1/k^{n}).
\end{eqnarray}

As a result, under the approximation conditions of the extreme-relativistic
case, we are allowed to take the matrix form of $T_{(M)}$, defined by (\ref{5.5a}), to be
\begin{eqnarray}
T_{(M)}(k)&=&\left(\begin{array}{cc}
t_{(M)}(k)&0\\
0&
\left(\begin{array}{cc}
1&0\\0&1
\end{array}\right)
\end{array}\right)\quad,\quad
t_{(M)}(k)=\left(\begin{array}{cc}
y^{(1)}_{1}(k)&y^{(2)}_{1}(k)\\
y^{(1)}_{2}(k)&y^{(2)}_{2}(k)
\end{array}\right),\\
\frac{y^{(1)}_{1}(k)}{y^{(1)}_{2}(k)}&=&\frac{-A\sin(2\theta)}{-\Delta m^{2}
+A\cos(2\theta)+\sqrt{(A\cos(2\theta)-\Delta m^{2})^{2}+(A\sin(2\theta))^{2}}},
\label{5.17c}\\
\frac{y^{(2)}_{1}(k)}{y^{(2)}_{2}(k)}&=&\frac{-A\sin(2\theta)}{-\Delta m^{2}
+A\cos(2\theta)-\sqrt{(A\cos(2\theta)-\Delta m^{2})^{2}+(A\sin(2\theta))^{2}}}
\label{5.17d}
\end{eqnarray}
with the unitarity condition $t^{\dagger}_{(M)}t_{(M)}=I$.

Thus the relation of $\{{\mib\alpha}_{M}(k\downarrow;x^{0}),
{\mib\beta}^{\dagger}_{M}(-k\downarrow;x^{0})\}$ with  $\{\tilde{\mib\alpha}_{M}
(k\downarrow;x^{0}),\tilde{\mib\beta}^{\dagger}_{M}(-k\downarrow;x^{0})\}$,
given by (\ref{5.8c}), becomes rather simple:
\begin{eqnarray}
&&{\mib\alpha}_{M}(k\downarrow;x^{0})=t_{M}(k)\tilde{\mib\alpha}_{M}
(k\downarrow;x^{0}),\\
&&{\mib\beta}^{\dagger}_{M}(-k\downarrow;x^{0})=
\tilde{\mib\beta}^{\dagger}_{M}(-k\downarrow;x^{0}).
\end{eqnarray}
On the other hand, the corresponding relation for $\{{\mib\alpha}_{F},{\mib\beta}_{F}\}$
includes the Bogolyubov transformation:
\begin{eqnarray}\label{5.19}
\left(\begin{array}{c}
{\mib\alpha}_{F}(k\downarrow;x^{0})\\{\mib\beta}^{\dagger}_{F}(-k\downarrow;x^{0})
\end{array}\right)={\cal K}(k)
\left(\begin{array}{c}
t_{M}(k)\tilde{\mib\alpha}_{M}(k\downarrow;x^{0})\\
\tilde{\mib\beta}^{\dagger}_{M}(-k\downarrow;x^{0})
\end{array}\right).
\end{eqnarray}
Under the approximation of the extreme-relativistic case, the {\it off-diagonal}
submatrix of ${\cal K}(k)$, i.e. $i\Lambda=i[z^{1/2}_{\rho j}\lambda _{\rho j}(k)]$
can not be set equal to zero, because  $\lambda _{\rho j}(k)=\sin((\chi _{\rho}-\chi _{j})/2)$
depends on $\{\mu_{\lambda}\}$;
in other words, since (\ref{2.33}) has no logical basis, we are not allowed to take (\ref{2.34b}).
Therefore, contrary to Ref.\cite{MAN}, we can not derive the
standard relations including the oscillation formulas explained in Sec.II.C.

Finally, we add a remark on Ref.\cite{MAN}.
When we write $t_{(M)}(k)$ as
\begin{eqnarray}
t_{M}(k)=\left(\begin{array}{cc}
\cos\theta _{M}&\sin\theta _{M}\\-\sin\theta _{M}&\cos\theta _{M}
\end{array}\right)
\end{eqnarray}
we have from (\ref{5.14b}) (or from (\ref{5.17c}) and (\ref{5.17d}))
\begin{eqnarray}
\tan\theta _{M}=\frac{m^{2}_{1}-\tilde m_{(+)1}(\downarrow)^{2}+A_{0}+
A\cos^{2}\theta_{\theta}}{A\cos\theta \sin\theta}.
\end{eqnarray}
This is equal to Eq.(2.25) in Ref.\cite{MAN} for $m^{2}_{A}=\tilde m_{(+)1}(\downarrow)^{2}$ and $A_{0}=0$.
As noted in Sec.II.C, the usually employed relations are
derived if we take the super-relativistic limit (\ref{2.34a}) and (\ref{2.34b}).
In this limit, ${\cal K}(k)$ reduces to
\begin{eqnarray}\label{5.21a}
{\cal K}(k)\longrightarrow
\left(\begin{array}{cc}
Z^{1/2}&0\\0&Z^{1/2}
\end{array}\right);
\end{eqnarray}
(\ref{5.19}) leads to
\begin{eqnarray}\label{5.21b}
{\mib\alpha}_{F}(k\downarrow;x^{0})=Z^{1/2}
t_{M}(k)\tilde{\mib\alpha}_{M}(k\downarrow;x^{0})
\end{eqnarray}
with
\begin{eqnarray}\label{5.21c}
Z^{1/2}t_{M}(k)=\left(\begin{array}{cc}
\cos(\theta+\theta _{M})&\sin(\theta+\theta _{M})\\
-\sin(\theta+\theta _{M})&\cos(\theta+\theta _{M});
\end{array}\right)
\end{eqnarray}
(\ref{5.21c}) should be equal to (\ref{2.38c}), i.e.
\begin{eqnarray}\label{5.21d}
Z^{1/2}t_{M}(k)=\left(\begin{array}{cc}
\cos\tilde\theta&\sin\tilde\theta\\
-\sin\tilde\theta&\cos\tilde\theta
\end{array}\right).
\end{eqnarray}
By employing (\ref{5.17c})$=-1/\tan\theta_{M}$ (or (\ref{5.17d})
$=\tan\theta_{M}$) as well as $\tan\tilde\theta$ given by (\ref{2.39a}),
we can confirm the equality (\ref{5.21d}) to hold.
As to the oscillation formulas,
we obtain (\ref{2.42b}) and(\ref{2.42c});
by using (\ref{2.39b}), (\ref{2.42b}) is rewritten as
\begin{eqnarray}
P^{matt}(\nu _{e}\rightarrow\nu _{\mu})=\frac{(\Delta m^{2}
\sin(2\theta))^{2}\sin^{2}(\Delta E^{matt}L/2)}{(A-\Delta m^{2}
\cos(2\theta))^{2}+(\Delta m^{2}\sin(2\theta))^{2}},
\end{eqnarray}
which is equal to  Eq.(2.35) in Ref.\cite{MAN}.

It is asserted in Ref.\cite{MAN} that the standard formalism of
neutrino oscillation in matter is derived from the field theoretical approach.
This assertion, however, is based on the approximation (\ref{5.21a})
which is not allowed when the relativistic approximation is consistently performed.

\section{conclusion and final remarks}\label{conclusions}
We have investigated in detail the  connection of the
operatorial field mixing relations
$\nu _{\rho}(x)=\sum_{j}z^{1/2}_{\rho j}\nu _{j}(x)$
with the usually employed formulas of neutrino oscillations in matter.
To this aim, we first derived the energy-eigenvalue equations
for the 2-flavor neutrinos in the static medium with the
$(x^{\rho})$-independent electron density.

We examined the high-energy neutrino case, called
{\it the extreme-relativistic case}, defined by (\ref{2.23}),
where all terms with higher order than $\{m^{2}_{j}, A, A_{0}\}/{k^2}$
in comparison with 1 are neglected and corrections to terms such
as $m^{2}_{j}$ and $k\kappa _{jl}$ are dropped.
After performing consistently this approximation, we obtained the
energy eigenvalues which coincide with the usual ones obtained in
the quantum mechanical approach\cite{FY} as well as in the
field theoretical one by Mannheim\cite{MAN}.

As for the evolution relations, however, we obtained a negative
result against  Mannheim's assertion that the standard formalism
of the neutrino oscillation is derived in the field theoretical approach.
The reason for such a difference has been pointed out
in Sec.V:
while the relation (\ref{5.19}) in our case, i.e.
\begin{eqnarray}\label{6.1}
\alpha _{\rho}(k\downarrow;x^{0})=\sum _{j}\lf\{{\cal K}(k)_{\rho j}
(t_{(M)}\tilde\alpha _{M}(k\downarrow;x^{0}))_{j}+{\cal K}(k)_{\rho \bar j}
\tilde\beta^{\dagger}_{j}(-k\downarrow;x^{0})\ri\},
\end{eqnarray}
is (consistently) obtained,
in Mannheim's case\cite{MAN} the relation (\ref{5.21b}), i.e.
\begin{eqnarray}\label{6.2}
\alpha _{\rho}(k\downarrow;x^{0})=\sum _{j}z^{1/2}_{\rho j}(t_{(M)}
\tilde\alpha _{M}(k\downarrow;x^{0}))_{j},
\end{eqnarray}
is utilized.
Eq.(\ref{6.2}) is obtained from (\ref{6.1}) by performing (\ref{2.34b})
\begin{eqnarray}\label{6.3}
\rho _{\rho j}\longrightarrow 1\quad, \quad
\lambda _{\rho j}\longrightarrow 0\quad, \qquad
\forall\rho,j;
\end{eqnarray}
then the Bogolyubov transformation (\ref{2.32b}) reduces to
\begin{eqnarray}
\left(\begin{array}{c}
{\mib\alpha}_{F}(kr;x^{0})\\{\mib\beta}^{\dagger}_{F}(-kr;x^{0})
\end{array}\right)
=\left(\begin{array}{cc}
z^{1/2}&0\\0&z^{1/2}
\end{array}\right)
\left(\begin{array}{c}
{\mib\alpha}_{M}(kr;x^{0})\\{\mib\beta}^{\dagger}_{M}(-kr;x^{0})
\end{array}\right)
\end{eqnarray}
where the most characteristic feature of the field theory of mixed fields
\cite{BV1,FHY1,BV2} has been lost.
As already stressed in Sec.V, $\rho _{\rho j}$ and $\lambda _{\rho j}$
depend on the arbitrary parameters $\{\mu _{\lambda}\}$, and the limit
(\ref{6.3}) has no connection with the approximation in
{\it the extreme-relativistic case} (\ref{2.23}) with (\ref{2.24}).

When the electron density in medium $n_{(e)}$ is taken to be $0$, i.e.
$A_{0}\rightarrow 0, A\rightarrow 0$, various relations in medium reduce to those in vacuum.
It is
worth to remember that, although our considerations have been
done for the 2-flavor case, the results obtained have qualitatively
general features which hold true for the
many flavors case both in matter and in vacuum.

Regarding the calculation of neutrino oscillations in quantum field theory,
there are several possible approaches \cite{GKL,GS,CARD,YI},
in which transition amplitudes for a finite time interval or finite distance are
calculated on the basis of the field theory by treating the neutrinos as
particles in intermediate states and without constructing any one-particle
state with definite flavor.
In the line  of the considerations developed in the present paper, it seems
meaningful for us to investigate  further
the implications of the relation (\ref{5.12a}) .
This is a kind of  evolution relation corresponding to the usual one
given by (\ref{2.40b}), and the matrix $W^{(F)}(k\downarrow;x^{0})$,
corresponding to the usual transition amplitudes (\ref{2.42a}), is an
extension of $W(k;x^{0})$ introduced in Ref.\cite{FHY2} in the vacuum case.
It has been pointed out\cite{FHY2} that
\begin{eqnarray}
W(k;x^{0})=
\left(\begin{array}{cc}
(W_{\rho\sigma}(k;x^{0}))&(W_{\rho\bar\sigma}(k;x^{0}))\\(W_{\bar\rho\sigma}
(k;x^{0}))&(W_{\bar\rho\bar\sigma}(k;x^{0}))
\end{array}\right)
\end{eqnarray}
has an intimate connection with the momentum components of the retarded
propagators or the anticommutators of the flavor neutrino fields as well as with the
expectation value of the flavor charges\cite{BHV}. The combinations
\begin{eqnarray}
\begin{array}{c}
\frac{1}{2}\left\{ |(W_{\rho\sigma}(k;x^{0}))|^{2}+|(W_{\rho\bar\sigma}(k;x^{0}))|^{2}
+|(W_{\bar\rho\sigma}(k;x^{0}))|^{2}+|(W_{\bar\rho\bar\sigma}(k;x^{0}))|^{2}\right\}
\end{array}
\end{eqnarray}
are independent of $\{\mu _{\lambda}\}$ for $\forall\rho,\sigma$, and have
properties allowing a probability interpretation.
$W^{(F)}(k\downarrow;x^{0})$ is also seen to satisfy the above properties.
In a subsequent
paper\cite{FH} we will investigate a possible field theoretical approach to the neutrino oscillation, where possible implications of the matrices $W(k;x^{0})$
and $W^{(F)}(k\downarrow;x^{0})$  to the oscillation formula are considered.

\section*{Acknowledgements}
The authors would like to express their thanks to Prof. G.Vitiello, the
organizer of International  Workshop on the Quantum Field Theory of Particle
mixing and Oscillations, 13-15 June 2002, Vietri sul Mare (Salerno),
 where the main content of the present paper was report, and also
to the participants in that workshop for discussions, especially
to Dr. C.Y.Cardall for his calling attention to the Mannheim's
paper. M.B. thanks the ESF network COSLAB and  EPSRC for support.
\appendix
\section{Momentum-helicity eigenfunctions}
Explicit forms of the momentum-helicity eigenfunctions satisfying
\begin{eqnarray}\label{A1}
(i\not k+m)u(kr)=0\,,\quad  (-i\not k+m)v(kr)=0
\end{eqnarray}
are given in the Kramers representation ($\vec\gamma=-\rho _{y}
\otimes\vec\sigma, \gamma^{4}=\rho _{x}\otimes I,\,\,  \gamma _{5}=
\gamma^{1}\gamma^{2}\gamma^{3}\gamma^{4}=-\rho _{z}\otimes I$)\cite{CORN} by
\begin{eqnarray}
u(k\uparrow)=\left(\begin{array}{c}
c\alpha\\ c\beta\\ s\alpha\\ s\beta
\end{array}\right),\;
u(k\downarrow)=\left(\begin{array}{c}
-s\beta^{\ast}\\ s\alpha^{\ast}\\ -c\beta^{\ast}\\ c\alpha^{\ast}
\end{array}\right),\;
v(k\uparrow)=\left(\begin{array}{c}
s\beta^{\ast}\\ -s\alpha^{\ast}\\ -c\beta^{\ast}\\ c\alpha^{\ast}
\end{array}\right),\;
v(k\downarrow)=\left(\begin{array}{c}
c\alpha\\ c\beta\\ -s\alpha\\ -s\beta
\end{array}\right),
\end{eqnarray}
where
\begin{eqnarray}
c=\cos(\chi/2)\quad,\quad &&s=\sin(\chi/2)\quad,\quad
 cot\chi=|\vec k|/m, k_{z}=|\vec k|\cos\vartheta
 \quad,\quad
k_{x}+ik_{y}=|\vec k|\sin\vartheta e^{i\phi}, \nonumber\\
&&\alpha=\cos(\vartheta/2)e^{-i\phi/2}\quad,\quad
 \beta=\sin(\vartheta/2)e^{i\phi/2};
\end{eqnarray}
for $\vec s=(I\otimes\vec\sigma)/2$ and $\hat{\vec k}=\vec k/|\vec k|$, we have
\begin{eqnarray}
(\vec s\hat{\vec k})u(kr)=\pm\frac{1}{2}u(kr)\,,\quad (\vec s\hat{\vec k})v(-kr)=
\pm\frac{1}{2}v(-kr) \quad \mbox{ for }\quad r=\left\{\begin{array}{c}
\uparrow\\\downarrow
\end{array}\right\}.
\end{eqnarray}
Here, $v(-kr)$ is defined by
\begin{eqnarray}
v(-kr)\equiv v(-\vec k,k^{0},r).
\end{eqnarray}
Noting $\alpha(-\vec k)=-i\beta^{\ast}, \beta(-\vec k)=i\alpha^{\ast}$, we have
\begin{eqnarray}
v(-k\downarrow)=i\left(\begin{array}{c}
-s\alpha\\ -s\beta\\ c\alpha\\ c\beta
\end{array}\right)\,,\quad
v(-k\uparrow)=i\left(\begin{array}{c}
-c\beta^{\ast}\\ c\alpha^{\ast}\\ s\beta^{\ast}\\ -s\alpha^{\ast}
\end{array}\right).
\end{eqnarray}

We write the solution of Eq.(\ref{A1}) with mass $m$ as $u_{j}(kr)$ and $v_{j}(kr)$.
Then we have
\begin{eqnarray}
u^{\ast}_{j}(kr)u_{l}(ks)&=&v^{\ast}_{j}(-kr)v_{l}(-ks)=
\rho _{jl}(k)\delta _{rs},\nonumber\\
u^{\ast}_{j}(kr)v_{l}(-ks)&=&v^{\ast}_{j}(-kr)u_{l}(ks)=
i\lambda _{jl}(k)\delta _{rs},\nonumber\\
\end{eqnarray}
where $\rho _{jl}(k)=\cos[(\chi _{j}-\chi _{l})/2], \lambda _{jl}(k)
=\sin[(\chi _{j}-\chi _{l})/2]$ with $cot\chi _{j}=|\vec k|/m_{j}$.
Further we have
\begin{eqnarray}
\sum _{r}\{ u(kr)u^{\ast}(kr)+v(-kr)v^{\ast}(-kr)\}=I^{(4)}.
\end{eqnarray}
($I^{(4)}$ is the $4\times 4$ unit matrix.)

Utilizing the concrete forms of $u(kr)$ and $v(kr)$, we obtain Table I,
where $P_{L}=(1+\gamma _{5})/2,$
 $s_{j}\equiv \sin(\chi _{j}/2), c_{j}\equiv \cos(\chi _{j}/2), cot\chi _{j}=|\vec k|/m_{j}$.
Further, we obtain (by writing $2\vec s=\vec\sigma$)
\begin{eqnarray}\label{A9}
u^{\dagger}_{j}(kr)(\vec\sigma\vec u)P_{L}u_{l}(kr)&=&
u^{\dagger}_{j}(kr)\frac{1}{2}\{(\vec\sigma\vec u),(\vec\sigma\hat{\vec k})\}P_{L}u_{l}(kr)
\left\{\begin{array}{c}+1\\-1\end{array}\right\}
\nonumber\\
&=&(\vec u\hat{\vec k})u^{\dagger}_{j}(kr)P_{L}u_{l}(kr)
\left\{\begin{array}{c}+1\\-1\end{array}\right\}
\nonumber\\
&=&(\vec u\hat{\vec k})\left\{\begin{array}{c}+s_{j}s_{l}
\\-c_{j}c_{l}\end{array}\right\}
\mbox{ for }r=\left\{\begin{array}{c}+1\\-1\end{array}\right\},
\\
u^{\dagger}_{j}(k\uparrow)(\vec\sigma\vec u)P_{L}u_{l}(k\downarrow)
&=&\frac{1}{2}u^{\dagger}_{j}(k\uparrow)[(\vec\sigma\hat{\vec k}),
(\vec\sigma\vec u)]P_{L}u_{l}(k\downarrow)\nonumber\\
&=&i(\hat{\vec k}\times\vec u)u^{\dagger}_{j}(k\uparrow)\vec
\sigma P_{L}u_{l}(k\downarrow)\not\equiv 0,\label{A10}
\end{eqnarray}
and similar relations for other combinations among the momentum-helicity eigenfunctions.

\begin{center}
\begin{tabular}{|c|c|c|c|c|}
\hline
$(r,s)$&$(\uparrow,\uparrow)$&$(\downarrow,\downarrow)$&
$(\uparrow,\downarrow)$&$(\downarrow,\uparrow)$\\
\hline
$u^{\dagger}_{j}(kr)P_{L}u_{l}(ks)$&$s_{j}s_{l}$&$c_{j}c_{l}$&$0$&$0$\\
\hline
$v^{\dagger}_{j}(-kr)P_{L}v_{l}(-ks)$&$c_{j}c_{l}$&$s_{j}s_{l}$&$0$&$0$\\
\hline
$u^{\dagger}_{j}(kr)P_{L}v_{l}(-ks)$&$is_{j}c_{l}$&$-ic_{j}s_{l}$&$0$&$0$\\
\hline
$v^{\dagger}_{j}(-kr)P_{L}u_{l}(ks)$&$-ic_{j}s_{l}$&$is_{j}c_{l}$&$0$&$0$\\
\hline
\end{tabular}
\\ [2mm]
Table I.
\end{center}
\section{Eigenvalues of $H^{matt}_{(M)}$ in the extreme-relativistic case}\label{eigenvalues}
We examine the eigenvalues of $H^{matt}_{(M)}(kr)$, i.e. the solutions of
\begin{eqnarray}
\det[H^{matt}_{(M)}(kr)-\lambda]=0\quad , \quad r=\uparrow, \downarrow.
\end{eqnarray}
First we consider $H^{matt}_{(M)}(k\downarrow)$, because it has a
relation with ${\mib\alpha}_{F}(k\downarrow;x^{0})$-operators of the
neutrinos produced through the weak V-A interaction;
then, from (\ref{2.19b})(or (\ref{3.10c})) we examine the equation
\begin{eqnarray}
\label{4.2}
\det[H^{matt}_{(M)}(k\downarrow)-\lambda]&=&(\lambda^{2}-k^{2}-m^{2}_{1})
(\lambda^{2}-k^{2}-m^{2}_{2})\nonumber\\
&-&(\lambda+k)\lf\{\kappa _{11}(\lambda^{2}-k^{2}-m^{2}_{2})+\kappa _{22}
(\lambda^{2}-k^{2}-m^{2}_{1})\ri\}+(\lambda+k)^{2}\det\kappa=0.
\end{eqnarray}
We search its approximate solutions in the {\it extreme-relativistic case}
(\ref{2.23}) by taking into consideration all terms of $1/k^{2}$-order in comparison with $1$.
Thus, the lowest order solutions of (\ref{4.2}) are $\lambda =\pm k$, so
that we examine solutions in the neighborhoods of $k$ and $-k$ separately,
which we call hereafter $(+)$- and $(-)$-solutions, respectively.

\subsection{$(+)$-solutions}\label{solutions}
Eq.(\ref{4.2}) is rewritten as
\begin{eqnarray}
\det[H^{matt}_{(M)}(k\downarrow)-\lambda]&=&(\lambda+k)^{2}F^{\downarrow}_{(+)}(\lambda)=0,
\end{eqnarray}
where
\begin{eqnarray}
F^{\downarrow}_{(+)}(\lambda)&\equiv&(\lambda-k)^{2}-(\lambda-k)
\lf\{\frac{m^{2}_{1}+m^{2}_{2}}{\lambda+k}+(\kappa _{11}+\kappa _{22})\ri\}
+\lf\{\frac{m^{2}_{1}m^{2}_{2}}{(\lambda+k)^{2}}+\frac{\kappa _{11}m^{2}_{2}+
\kappa _{22}m^{2}_{1}}{\lambda+k}+\det\kappa\ri\}.
\end{eqnarray}
The discriminant $D_{(+)}$ of $F^{\downarrow}_{(+)}(\lambda)$,
regarded as a quadratic equation with respect to $(\lambda-k)$, is
\begin{eqnarray}
D_{(+)}=\lf\{\frac{\Delta m^{2}}{\lambda+k}-b_{1}\cos(2\theta)\ri\}^{2}
+b^{2}_{1}\sin^{2}(2\theta);
\end{eqnarray}
thus, the $(+)$-solutions satisfy
\begin{eqnarray}
\lambda _{(+)}-k=\frac{1}{2}\lf(\frac{m^{2}_{1}+m^{2}_{2}}{\lambda _{(+)}+k}+
2b_{0}+b_{1}\ri)\pm\frac{1}{2}\sqrt{D_{(+)}(\lambda _{(+)})}.\label{4.4b}
\end{eqnarray}
With the aim of comparing (\ref{4.4b}) with (\ref{2.37a}), we define
\begin{eqnarray}
\lambda _{(+)+}-k=\frac{\tilde m_{(+)2}(\downarrow)^{2}}{2k}\quad, \quad
\lambda _{(+)-}-k=\frac{\tilde m_{(+)1}(\downarrow)^{2}}{2k};
\end{eqnarray}
here, $\lambda _{(+)+}$ and $\lambda _{(+)-}$ correspond to the solutions
with the signs $+$ and $-$, respectively, in front of $\sqrt{D_{(+)}}$ in  (\ref{4.4b}).
Then we have
\begin{eqnarray}
\tilde m_{(+)2}(\downarrow)^{2}&=&\frac{{\bar m^{2}}}{1+\tilde m_{(+)2}
(\downarrow)^{2}/(4k^{2})}+A_{0}+\frac{1}{2}[A+\sqrt{D_{(+)}(\lambda _{(+)+})(2k)^{2}}],
\\
\tilde m_{(+)1}(\downarrow)^{2}&=&\frac{{\bar m^{2}}}{1+\tilde m_{(+)1}
(\downarrow)^{2}/(4k^{2})}+A_{0}+\frac{1}{2}[A-\sqrt{D_{(+)}(\lambda _{(+)-})(2k)^{2}}].
\end{eqnarray}
In accordance with the explanation concerning the approximation given in
Sec.II.B, we have the approximate forms of
$D_{(+)}(\lambda _{(+)\pm})$ expressed as
\begin{eqnarray}
D_{(+)}(\lambda _{(+)+})(2k)^{2}&=&\{\frac{\Delta m^{2}}{1+\tilde m_{(+)2}(\downarrow)^{2}
/(4k^{2})}-A \cos(2\theta)\}^{2}+\{A\sin(2\theta)\}^{2};\nonumber
\\
&\simeq&\{\Delta m^{2}-A \cos(2\theta)\}^{2}+\{A \sin(2\theta)\}^{2},
\\
D_{(+)}(\lambda _{(+)-})(2k)^{2}&\simeq&\{\Delta m^{2}-A
\cos(2\theta)\}^{2}+\{A \sin(2\theta)\}^{2};
\end{eqnarray}
thus we have
\begin{eqnarray}\label{4.7}
\left(\begin{array}{c}
\tilde m_{(+)1}(\downarrow)^{2}\\\tilde m_{(+)2}(\downarrow)^{2}
\end{array}\right)
\simeq {\bar m^{2}}+A_{0}+\frac{1}{2}\{A+\sqrt{(A-\Delta m^{2}\cos(2\theta))^{2}
+(\Delta m^{2}\sin(2\theta))^{2}}
\left(\begin{array}{c}
-1\\+1
\end{array}\right)\},
\end{eqnarray}
which coincide with the usually utilized expressions
$\tilde m_{j}(\downarrow)^{2}$'s given by (\ref{2.37b}).

\subsection{$(-)$-solutions}\label{solutions2}
Eq.(\ref{4.2}) is rewritten as
\begin{eqnarray}
&&\det[H^{matt}_{(M)}(k\downarrow)-\lambda]\,=\,(\lambda-k)^{2}
F^{\downarrow}_{(-)}(\lambda)=0,\\
&&F^{\downarrow}_{(-)}(\lambda)\,\equiv\,(\lambda+k)^{2}\{1-
\frac{\kappa _{11}+\kappa _{22}}{\lambda-k}+\frac{\det\kappa}{(\lambda-k)^{2}}\}
+(\lambda+k)\lf\{- \frac{m^{2}_{1}+m^{2}_{2}}{\lambda-k}+
\frac{\kappa _{11}m^{2}_{2}+\kappa _{22}m^{2}_{1}}{(\lambda-k)^{2}}\ri\}
+\frac{m^{2}_{1}m^{2}_{2}}{(\lambda-k)^{2}},
\end{eqnarray}
\begin{eqnarray}
\mbox{Discriminant }D_{(-)}&=&\lf\{- \frac{m^{2}_{1}+m^{2}_{2}}{\lambda-k}
+\frac{\kappa _{11}m^{2}_{2}+\kappa _{22}m^{2}_{1}}{(\lambda-k)^{2}}\ri\}^{2}
-\frac{4m^{2}_{1}m^{2}_{2}}{(\lambda-k)^{2}}\lf\{1-\frac{\kappa _{11}
+\kappa _{22}}{\lambda-k}+\frac{\det\kappa}{(\lambda-k)^{2}}\ri\}
\nonumber\\
&=&\frac{1}{(\lambda-k)^{2}}\lf[\lf\{\Delta m^{2}-\frac{\kappa _{11}m^{2}_{2}
-\kappa _{22}m^{2}_{1}}{\lambda-k}\ri\}^{2}+
\frac{4m^{2}_{1}m^{2}_{2}}{(\lambda-k)^{2}}\kappa^{2}_{12}\ri];
\end{eqnarray}
thus, the $(-)$-solutions satisfy
\begin{eqnarray}
\lambda _{(-)}+k=\frac{1}{2(1-\frac{b_{0}}{\lambda _{(-)}-k})
(1-\frac{b_{0}+b_{1}}{\lambda _{(-)}-k})}
\left[ \frac{2{\bar m^{2}}}{\lambda _{(-)}-k}-\frac{(2b_{0}+b_{1})
{\bar m^{2}}+b_{1}\cos(2\theta)\Delta m^{2}/2}{(\lambda _{(-)}-k)^{2}}
\mp\sqrt{D_{(-)}(\lambda _{(-)})}\right].
\end{eqnarray}
We define $\tilde m_{j}(\downarrow)^{2}, j=1,2$, as
\begin{eqnarray}
\lambda _{(-)-}+k=-\frac{\tilde m_{(-)2}(\downarrow)^{2}}{2k}\quad , \quad
\,\lambda _{(-)+}+k=-\frac{\tilde m_{(-)1}(\downarrow)^{2}}{2k}.
\end{eqnarray}
In the same approximation as before, we have
\begin{eqnarray}
\sqrt{D_{(-)}(\lambda _{(-)\mp})(2k)^{2}}\simeq \sqrt{D_{(-)}(-k)(2k)^{2}}\simeq\Delta m^{2};
\end{eqnarray}
thus,
\begin{eqnarray}
\tilde m_{(-)j}(\downarrow)^{2}\simeq m_{j}^{2},\quad  j=1,2.
\end{eqnarray}
\subsection{remarks}
Similarly to the $H^{matt}_{(M)}(k\downarrow)$ case, we obtain the
 approximate eigenvalues of $H^{matt}_{(M)}(k\uparrow)$.
We use the notations $\lambda _{(+)j}$ and $\lambda _{(-)j}$($j=1,2$)
for expressing the solutions of $\det[H^{matt}_{(M)}(k\uparrow)-0]=0$.
By comparing (\ref{2.17c}) and (\ref{2.19a}) with (\ref{2.17d}) and
(\ref{2.19b}), respectively, we see the approximate eigenvalues are given by
\begin{eqnarray}
\lambda _{(+)j}=k+\frac{\tilde m_{(+)j}(\uparrow)^{2}}{2k}\quad,\quad
 \lambda_{(-)j}=-k-\frac{\tilde m_{(-)j}(\uparrow)^{2}}{2k},
\end{eqnarray}
where
\begin{eqnarray}
&&\tilde m_{(+)j}(\uparrow)^{2}\simeq m_{j}^{2},\nonumber\\
&&\left(\begin{array}{c}
\tilde m_{(-)1}(\uparrow)^{2}\\\tilde m_{(-)2}(\uparrow)^{2}
\end{array}\right)
\simeq {\bar m^{2}}-A_{0}-\frac{1}{2}\{A+\sqrt{(A+\Delta m^{2}
\cos(2\theta))^{2}+(\Delta m^{2}\sin(2\theta))^{2}}
\left(\begin{array}{c}
+1\\-1
\end{array}\right)\}.
\end{eqnarray}

As seen from the considerations described above in the present section,
we obtain in the {\it extreme-relativistic case} the eigenvalues (\ref{4.7})
which coincide with the usually employed ones such as Eq.(\ref{2.37b}).
This does not necessarily mean, as stressed in Sec.II.C, the
validity of the usual relations among eigenvectors given below Eq.(\ref{2.37b}).
In Sec.IV, this point is examined in detail.

Notice that an analogous situation is described in Ref.\cite{BHV},
where the neutrino propagators on the flavor vacuum were studied:
although they share the same poles of the usual propagators
defined on the mass vacuum, they correspond to different boundary
conditions and lead to different oscillation amplitudes.


\end{document}